\documentclass[preprint,12pt]{elsarticle}

\usepackage{amsmath}
\usepackage{amssymb}
\usepackage{graphicx}% Include figure files
\usepackage{dcolumn}% Align table columns on decimal point
\usepackage{bm}% bold math
\begin{document}

\title{Driven - Dissipative Dynamics of Ultracold Atoms Trapped in an Array of Harmonic Potentials}% Force line breaks with \\

\author[ust,rcnas]{Roland Cristopher F. Caballar}
 %\altaffiliation[Also at ]{Department of Physics and Mathematics, College of Science, University of Santo Tomas}%Lines break automatically or can be forced with \\
\ead{rfcaballar@ust.edu.ph}
\affiliation[ust]{organization={Department of Mathematics and Physics, College of Science, University of Santo Tomas},
             addressline={Espana Boulevard, Sampaloc},
             city={Manila},
             postcode={1008},
             state={},
             country={Philippines}}
\affiliation[rcnas]{organization={Research Center for the Natural and Applied Sciences, University of Santo Tomas},
             addressline={Espana Boulevard, Sampaloc},
             city={Manila},
             postcode={1008},
             state={},
             country={Philippines}}
%\affiliation{%
% Department of Physics and Mathematics, College of Science, University of Santo Tomas
%
%\affiliation{Research Center for the Natural and Applied Sciences, University of Santo Tomas, Espana Blvd., Sampaloc, Manila, Philippines}


%\affiliation{
%\affiliation{
 %Third institution, the second for Charlie Author
%\affiliation{%


\date{\today}% It is always \today, today,

\begin{abstract}
We investigate the dynamics of a gas of ultracold atoms that are trapped in an array of harmonic potentials and that interacts with a Bose - Einstein condensate (BEC) that acts as a reservoir of Bogoliubov excitations. The ground and excited energy levels of these trapped ultracold atoms are coupled to each other via detuned Raman lasers with corresponding Rabi frequencies. Once excited via the Raman lasers, these trapped ultracold atoms then return to their ground energy levels, but not necessarily to their original trap locations, by emitting Bogoliubov excitations into the BEC. This combination of driving via Raman lasers to excited energy levels and dissipation via interaction with the BEC resulting in emission of Bogoliubov excitations into it will result in the trapped ultracold atoms approaching a steady state, whereby the expectation value of the number of trapped ultracold atoms in each harmonic trap of the array will attain a constant value over time. One can then use this system to prepare states that require a definite number of atoms in a particular energy level, such as BECs and atom lasers used for atom interferometry and for tests of the foundations of quantum mechanics.
%\begin{description}
%\end{description}
\end{abstract}

\maketitle

\section{Introduction}

In recent years, there has been an increase in the study and application of open quantum systems in quantum information and for quantum technologies \cite{BreuerPetruccione, breuer, devega}. Advances in optics and in atom trapping and cooling in particular have made it possible to create quantum systems that use the environment with which they interact as resources that drive their time evolution towards a particular quantum state \cite{verstraete,kastoryano2,kordas,reiter,nosov}. A particular mechanism that enables this is by using what is known as driven - dissipative dynamics, wherein a quantum system is driven from a lower energy level to a higher energy state, then decays back to its original energy level by interacting with an environment that acts as a reservoir of excitations that carry away energy from the system as it returns to its initial state. This continuous driving towards a higher energy level and dissipation of excitations to return to its original energy state will then cause the quantum system, coupled to an environment, to evolve towards a given steady - state, which is dependent on the choice of system, environment and interaction between the two. Driven - dissipative dynamics is widely used in dissipative quantum state preparation, wherein an initial quantum state evolves towards a desired final state by an appropriate choice of environment and interaction with this chosen environment. Examples of such dissipative quantum state preparation schemes are given in Refs. \cite{kastoryano,watanabe,stannigel,gong,sweke1,sweke2,su,shao,li,kouzelis,zheng,cole,sharma,seetharam,marino,yang}, which make use of various systems and various environments to prepare quantum states of interest in quantum information and quantum computing. Of particular interest among dissipative quantum state preparation schemes are those that make use of driven - dissipative dynamics to prepare of Bose - Einstein condensates (BECs), such as those described in Refs. \cite{vorberg,zhang} due to their widespread use and application in quantum computation and fundamental tests of quantum mechanics \cite{PethickSmith,PitaevskiiStringari}. One particular dissipative BEC preparation scheme is detailed in Refs. \cite{caspar1,caspar2}, wherein a universal quantum dissipative process is formulated using an appropriate choice of jump operators for a Markovian master equation describing the dissipative dynamics of spin - 1/2 bosons on a spatial lattice, with the system evolving via this master equation towards a BEC of hard-core spin - 1/2 bosons. Another dissipative BEC preparation scheme was described in Ref. \cite{walker}, wherein it was experimentally shown that driven - dissipative dynamics can be used in a dye - filled microcavity with light leaking out on both sides to create a photon BEC, albeit one with a small population (roughly of order of magnitude 1) of photons. 

In most dissipative quantum state preparation schemes, the environment is either a thermal bath that absorbs thermal excitations emitted by trapped atoms, or an optical bath that absorbs photons emitted from leaky optical cavities. However, in Ref. \cite{daley}, a new type of dissipative quantum preparation scheme for encoding qubits was formulated, one which made use of a system of trapped ultracold atoms immersed in a superfluid bath, which would play the role of a reservoir of Bogoliubov excitations emitted by the trapped ultracold atoms as they are driven from their initial energy states to excited energy states. This dissipative quantum state preparation scheme was later adapted in Ref. \cite{diehl} for a gas of ultracold atoms trapped in an optical lattice and immersed in a bath of Bogoliubov excitations, with the trapped ultracold atoms evolving over time towards a many-body pure state via the driven - dissipative dynamics of this quantum state preparation scheme. This scheme was also adapted and further refined in Refs. \cite{caballar,caballar2}, wherein a gas of ultracold atoms was trapped in a double well potential immersed in a harmonic trap and coupled to a background BEC acting as a reservoir of Bogoliubov excitations, with the driven - dissipative dynamics causing the trapped ultracold atoms to evolve over time towards a many - body phase or particle number squeezed state (which was done in Ref. \cite{caballar}) or a many - body spin steady state (as was done in Ref. \cite{caballar2}). 

Having shown that it is possible to formulate a dissipative quantum preparation scheme which uses a background BEC as a reservoir of Bogoliubov excitations which is coupled to a system of trapped ultracold atoms, we then investigate in this paper whether it is possible to use a variation of this scheme to prepare a many - body quantum state whose particle number expectation value remains constant over time, i. e. a trapped ultracold atom gas with a constant particle number expectation value. In doing so, the resulting dissipative quantum state preparation scheme can then be used as a basis for formulating a dissipative BEC preparation scheme, since one of the requirements for a BEC is that almost all, if not all, of the bosons comprising the condensate occupy the same energy level, thus providing an alternative to the dissipative BEC preparation schemes mentioned in Refs. \cite{vorberg,caspar1,caspar2,walker,zhang}. To do that, we first describe, in section 2 of this paper, the trapped ultracold atom system and the background BEC that we will be using for our scheme, together with the manner of interaction between system and background BEC, expressed in terms of the system, background BEC and interaction Hamiltonians, respectively. In particular, the trapping potential to be used for our ultracold atom system is an array of harmonic potentials, wherein three of these harmonic potentials will be loaded with atoms from our ultracold atom gas system, and will occupy the ground state energy levels of this trap. These atoms occupying the ground state energy levels in these harmonic potentials will then be driven, using Raman lasers with corresponding Rabi frequencies, to excited energy levels in the adjacent harmonic potential traps in the array. Once we have done this, we will then, in section 3, derive the master equation that will describe the driven - dissipative dynamics of our trapped ultracold atom system while it interacts with the background BEC. In deriving the master equation, we will make use of certain approximations, such as the phonon approximation for the Bogoliubov excitations emitted by the trapped ultracold atom gas into the background BEC, and the Born - Markov approximation for the coupling between the system and the background BEC. Once we have obtained the master equation, we present, in section 4, the numerical results of our time evolution of various initial states of the trapped ultracold atom gas of the system using this derived master equation. Our results show that, for particular values of physical parameters describing both the trapped ultracold atom system and the background BEC, such as the coupling strength between the system and the background BEC, the expectation value of the number of atoms initially located in each of the harmonic potentials in the trap array, and the particular form of the initial state of the trapped ultracold atom gas, the trapped ultracold atom gas will, via the driven - dissipative dynamics resulting from its interaction with the background BEC as well as the continued excitation of the atoms in the ground state to the excited state, evolve towards a state where the expectation value of the particle number in each of the harmonic potentials in the trap array will remain constant over time. 

\section{Derivation of the Interaction Hamiltonian for the System}

For our trapped ultracold atom system, we assume that the trapping array consists of five adjacent harmonic oscillator potentials, whose schematic diagram is shown in figure \ref{fig:harmonictraparray} below.
\begin{figure}[htb]
\includegraphics[width=1.0\columnwidth, height=0.2\textheight]{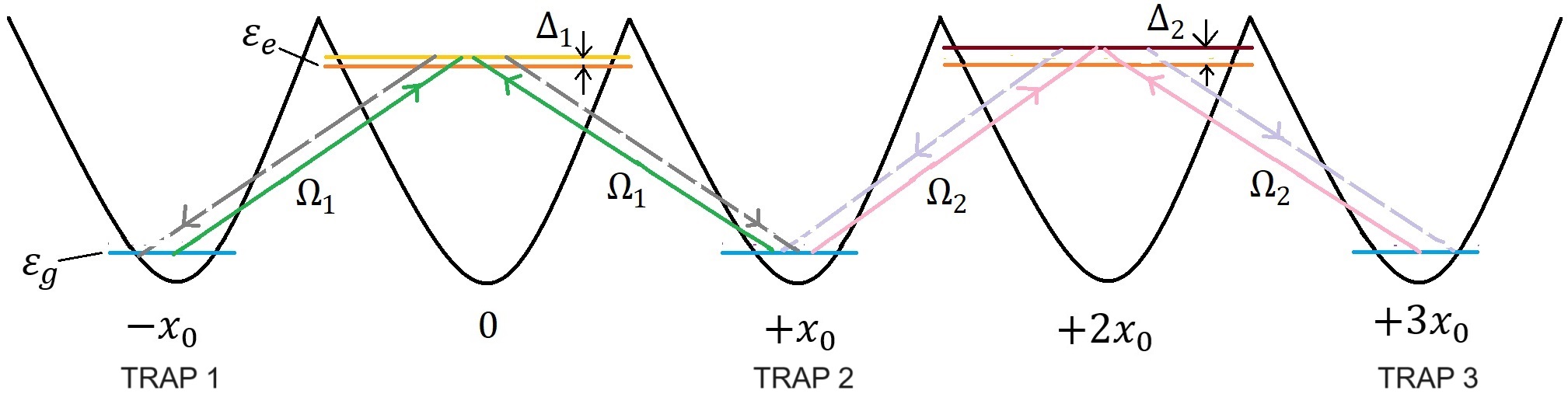}
\caption{\label{fig:harmonictraparray} Schematic diagram of the harmonic trap array and the driven - dissipative dynamics that occur as a result of the coupling between the ground state energy levels $\varepsilon_g$ (indicated by the blue lines) at traps 1, 2, and 3 (with centers located at $x=-x_0$, $x=+x_0$ and $x=+3x_0$, respectively) and the excited state energy levels $\varepsilon_e$ (indicated by the orange lines) at the harmonic traps with centers located at $x=0$ and $x=2x_0$, and the interaction between the trapped ultracold atoms and the background BEC. The coupling between energy levels is achieved by Raman lasers with Rabi frequencies $\Omega_1$ (green lines) for the ground state energy levels at traps 1 and 2, and $\Omega_2$ (pink lines) for the ground state energy levels at traps 2 and 3. These Raman lasers have corresponding detunings $\Delta_1$ (yellow line) and $\Delta_2$ (red line). On the other hand, as a result of the interaction between the trapped ultracold atoms and the background BEC, Bogoliubov excitations (indicated by dashed grey lines for atoms in traps 1 and 2 and by dashed lavender lines for atoms in traps 2 and 3) are emitted by atoms excited from the energy level $\varepsilon_g$ to the energy level $\varepsilon_e$ as they decay back to $\varepsilon_g$.}
\end{figure}

The corresponding Hamiltonian for the system can then be written as 

\begin{equation}
\hat{H}_{S}=\varepsilon_{g,1}\hat{a}^{\dagger}_{g,1}\hat{a}_{g,1}+\varepsilon_{g,2}\hat{a}^{\dagger}_{g,2}\hat{a}_{g,2}+\varepsilon_{g,3}\hat{a}^{\dagger}_{g,3}\hat{a}_{g,3}+\varepsilon_{e,1}\hat{a}^{\dagger}_{e,1}\hat{a}_{e,1}+\varepsilon_{e,2}\hat{a}^{\dagger}_{e,2}\hat{a}_{e,2}
\label{sysham}
\end{equation}

We note that the ground state energies in nodes 1, 2 and 3 are all equal to each other, i. e. $\varepsilon_{g,1}=\varepsilon_{g,2}=\varepsilon_{g,3}=\varepsilon_g$, and that the same is true for the excited state energies in the nodes between those corresponding to the ground state energies, i. e. $\varepsilon_{e,1}=\varepsilon_{e,2}=\varepsilon_e$. As shown in the figure, a pair of Rabi lasers with Raman frequencies $\Omega_1$ and detuning $\Delta_1$ are used to couple the ground states with energies $\varepsilon_{g,1}$ and $\varepsilon_{g,2}$ with the excited state with energy $\varepsilon_{e,1}$, while a second pair of Rabi lasers with Raman frequencies $\Omega_2$ and detuning $\Delta_2$ are used to couple the ground states with energies $\varepsilon_{g,2}$ and $\varepsilon_{g,3}$ with the excited state with energy $\varepsilon_{e,2}$. Hence, the trap sites corresponding to the excited states with energies $\varepsilon_{e,1}$ and $\varepsilon_{e,2}$ are coupled to the same site, which corresponds to the ground state with energy $\varepsilon_{g,2}$, in contrast to the other two sites corresponding to the ground states with energies $\varepsilon_{g,1}$ and $\varepsilon_{g,3}$, to which there is only one excited state coupled to each of these sites.

On the other hand, for the background BEC with which the trapped ultracold atom gas interacts, its Hamiltonian is given as
\begin{equation}
\hat{H}_{B}=\sum_{k}E_k \hat{b}^{\dagger}_{k}\hat{b}_k
\label{becham}
\end{equation}

One can consider the gas of ultracold atoms trapped in the array of harmonic potentials and the background BEC with which it interacts as a mixture of two ultracold atomic gases, for which the interaction Hamiltonian will have the following form:

\begin{equation}
\hat{H}_{SB}=\frac{2\pi a_{SB}}{\mu}\int dx\;\hat{\psi}^{\dagger}_{S}\hat{\psi}_{S}\hat{\psi}^{\dagger}_{B}\hat{\psi}_{B}
\label{interactham}
\end{equation}
Here, $\hat{\psi}_{S}$ and $\hat{\psi}_{B}$ are the field operators for the system of trapped ultracold atoms and the background BEC, respectively. For the system field operator, it can be written as

\begin{equation}
\hat{\psi}_{S}=\phi_{g,1}(x)\hat{a}_{g,1}+\phi_{g,2}(x)\hat{a}_{g,2}+\phi_{g,3}(x)\hat{a}_{g,3}+\psi_{e,1}(x)\hat{a}_{e,1}+\phi_{e,2}(x)\hat{a}_{e,2}
\end{equation}
where $\hat{a}_{n,j}$ and $\hat{a}^{\dagger}_{n,j}$ are the annihilation and creation operators, respectively, for the system corresponding to the energy level $\varepsilon_{n,j}$, which has corresponding energy eigenstates $\phi_{n,j}(x)$, with $n=g,e$ denoting the ground state or the excited state of the system, and $j=1,2,3$ denoting the location of the center of the energy eigenstate. In particular, for the ground state eigenstates,
\begin{eqnarray}
&&\phi_{g,1}(x)=\left(\frac{m_s \omega_g}{\pi\hbar}\right)^{1/4}\exp\left(-\frac{m_s \omega_g}{2\hbar}(x+x_0)^2\right)\underset{\sigma_g \rightarrow 0}\longrightarrow\sqrt{2}\pi^{1/4}\sigma_{g}^{1/2}\delta(x+x_0)\nonumber\\
&&\phi_{g,2}(x)=\left(\frac{m_s \omega_g}{\pi\hbar}\right)^{1/4}\exp\left(-\frac{m_s \omega_g}{2\hbar}(x-x_0)^2\right)\underset{\sigma_g \rightarrow 0}\longrightarrow\sqrt{2}\pi^{1/4}\sigma_{g}^{1/2}\delta(x-x_0)\nonumber\\
&&\phi_{g,3}(x)=\left(\frac{m_s \omega_g}{\pi\hbar}\right)^{1/4}\exp\left(-\frac{m_s \omega_g}{2\hbar}(x-3x_0)^2\right)\underset{\sigma_g \rightarrow 0}\longrightarrow\sqrt{2}\pi^{1/4}\sigma_{g}^{1/2}\delta(x-3x_0)\nonumber\\
\label{grndstateeigen}
\end{eqnarray}
wherein we make use of the property of Dirac delta distributions that
\begin{displaymath}
\frac{1}{\sqrt{2\pi}}\exp\left(-\frac{1}{\sigma^2}(x-x_0)^2\right)\underset{\sigma \rightarrow 0}\longrightarrow\delta(x-x_0)
\end{displaymath}
and where these eigenstates have width $\sigma_g = \sqrt{\frac{\hbar}{m_s \omega_g}}$. From these expressions, we see that the ground - state eigenstates $\phi_{g,1}(x)$, $\phi_{g,2}(x)$ and $\phi_{g,3}(x)$ of the system are centered at $x=-x_0$, $x=x_0$ and $x=3x_0$, respectively. On the other hand, for the excited state eigenstates,
\begin{eqnarray}
&&\phi_{e,1}(x)=\left(\frac{m_s \omega_e}{\pi\hbar}\right)^{1/4}\sqrt{\frac{2m_s \omega_e}{\hbar}}x\exp\left(-\frac{m_s \omega_e}{2\hbar}x^2\right)=\sqrt{2}\pi^{-1/4}\sigma_{e}^{-3/4}x\exp\left(-\frac{x^2}{2\sigma_e^2}\right)\nonumber\\
&&\phi_{e,2}(x)=\left(\frac{m_s \omega_e}{\pi\hbar}\right)^{1/4}\sqrt{\frac{2m_s \omega_e}{\hbar}}(x-2x_0 )\exp\left(-\frac{m_s \omega_e}{2\hbar}(x-2x_0 )^2\right)\nonumber\\
&&=\sqrt{2}\pi^{-1/4}\sigma_{e}^{-3/4}(x-2x_0 )\exp\left(-\frac{(x-2x_0 )^2}{2\sigma_e^2}\right)\nonumber\\
\label{exctdstateeigen}
\end{eqnarray}
These eigenstates have width $\sigma_e = \sqrt{\frac{\hbar}{m_s \omega_e}}$ and are centered at $x=0$ and $x=2x_0$, respectively. Multiplying the field operator of the system with its adjoint, and considering only terms up to first order in the system creation and annihilation operators $\hat{a}^{\dagger}_{n,j}$ and $\hat{a}_{n,j}$, we obtain

\begin{eqnarray}
&&\hat{\psi}^{\dagger}_{S}\hat{\psi}_S = \phi_{e,1}(x)\phi_{g,1}(x)(\hat{a}^{\dagger}_{e,1}\hat{a}_{g,1}+\hat{a}^{\dagger}_{g,1}\hat{a}_{e,1})+\phi_{e,1}(x)\phi_{g,2}(x)(\hat{a}^{\dagger}_{e,1}\hat{a}_{g,2}+\hat{a}^{\dagger}_{g,2}\hat{a}_{e,1})\nonumber\\
&&+\phi_{e,2}(x)\phi_{g,2}(x)(\hat{a}^{\dagger}_{e,2}\hat{a}_{g,2}+\hat{a}^{\dagger}_{g,2}\hat{a}_{e,2})+\phi_{e,2}(x)\phi_{g,3}(x)(\hat{a}^{\dagger}_{e,2}\hat{a}_{g,3}+\hat{a}^{\dagger}_{g,3}\hat{a}_{e,2})\nonumber\\
\label{sysfieldopprod}
\end{eqnarray}
We note that in computing for this product, we neglected the particle number terms $\hat{a}^{\dagger}_{n,j}\hat{a}_{n,j}$ as well as terms proportional to $\hat{a}^{\dagger}_{g,3}\hat{a}_{e,1}$ and its conjugate. The former is due to the assumption that particle number is conserved (and hence remains constant over time, so these terms have no contribution to the time evolution of the system), while the latter is due to the nearest - neighbor approximation, whereby atoms trapped in one site can only be coupled to energy levels in adjoining sites. This approximation allows us to avoid any complications that may arise due to the possibility of atoms tunneling between non-adjacent trap array sites. 

On the other hand, the field operator for the background BEC has the form

\begin{equation}
\hat{\psi}_{B}(x)=\sqrt{\rho_B}+\delta\hat{\psi}_B (x)
\end{equation}
Here, $\sqrt{\rho_B}$ is the BEC density, while $\delta\hat{\psi}_B$ is the condensate excitation term, and has the explicit form
\begin{equation}
\delta\hat{\psi}_B = \frac{1}{\sqrt{L}}\sum_{k}\left(u_k e^{ikx}\hat{b}_k +v_\mathbf{k}e^{-ikx}\hat{b}^{\dagger}_k \right),
\label{becexcifieldop}
\end{equation}
In this term, $L$ is the length of the BEC, $u_k =(1-L^{2}_k )^{-1/2}$, $v_k =L_k (1-L^{2}_k )^{-1/2}$, $L_k =(E_k -(k^2/2m_{B})-m_{B}c^2)/m_{B}c^2$, and $E_k$ is the excitation energy given by

\begin{equation}
E_k = ck\sqrt{1+\left(\frac{k}{2m_B c}\right)^2}
\end{equation}
with $c=\sqrt{g_{BB}\rho_B /m_B}$ being the velocity of sound in the condensate and $g_{BB}$ the interaction strength among the condensate atoms. Here, we make use of the assumption that the excitatons emitted into and from the BEC are sound - like, so that $E_k\approx ck$. However, this approximation implies that $k^2 / 4(m_B c)^2 <<1$. As such, the terms in the BEC field operator will then have the following form:
\begin{eqnarray}
&&L_k =\frac{1}{m_B c^2}\left(E_k -\frac{k^2}{2m_{B}}-m_{B}c^2\right)\approx \frac{k}{m_B c}-1\nonumber\\
&&u_k+v_k=\frac{1+L_{k}}{\sqrt{1-L^2_k}}=\sqrt{\frac{1+L_{k}}{1-L_{k}}}\approx\sqrt{\frac{k/m_B c}{2-k/m_B c}}\nonumber\\
&&=\sqrt{\frac{k}{2m_B c}}\left(1-\frac{k}{2m_B c}\right)^{-1/2}\approx\sqrt{\frac{k}{2m_B c}}\nonumber\\
\end{eqnarray}

The second of the two equations given above will be crucial in the derivation of the explicit form of $\hat{\psi}^{\dagger}_B\hat{\psi}_B$, which, if we are to keep only terms up to first order in $\hat{b}_k$ and $\hat{b}^{\dagger}_k$, can be written as follows:

\begin{eqnarray}
&&\hat{\psi}^{\dagger}_B\hat{\psi}_B = \left(\sqrt{\rho}_B+\frac{1}{\sqrt{L}}\sum_{k}(u_k e^{-ikx}\hat{b}^\dagger_k +v_k e^{ikx}\hat{b}_k )\right)\left(\sqrt{\rho}_B+\frac{1}{\sqrt{L}}\sum_{k}(u_k e^{ikx}\hat{b}_k + v_k e^{-ikx}\hat{b}^\dagger_k )\right)\nonumber\\
&&\approx\sqrt{\frac{\rho_B}{L}}\sum_{k}(u_k +v_k )(e^{ikx}\hat{b}_k+e^{-ikx}\hat{b}^\dagger_k )=\sqrt{\frac{\rho_B}{2m_B cV}}\sum_{k}\sqrt{k}(e^{ikx}\hat{b}_k +e^{-ikx}\hat{b}^\dagger_k )\nonumber\\
\label{becfieldopprod}
\end{eqnarray}

Thus, substituting Eqs. \ref{sysfieldopprod} and \ref{becfieldopprod} into Eq. \ref{interactham}, we obtain the following expression:

\begin{eqnarray}
&&\hat{H}_{SB}=\frac{2\pi a_{SB}}{\mu}\int dx\;\hat{\psi}^{\dagger}_{S}\hat{\psi}_{S}\hat{\psi}^{\dagger}_{B}\hat{\psi}_{B}\nonumber\\
&&=\frac{2\pi a_{SB}}{\mu}\sqrt{\frac{\rho_B}{2m_B cV}}\sum_{k}\sqrt{k}\int dx\;\left(\phi_{e,1}(x)\phi_{g,1}(x)(\hat{a}^{\dagger}_{e,1}\hat{a}_{g,1}+\hat{a}^{\dagger}_{g,1}\hat{a}_{e,1})+\phi_{e,1}(x)\phi_{g,2}(x)\right.\nonumber\\
&&\left.\times(\hat{a}^{\dagger}_{e,1}\hat{a}_{g,2}+\hat{a}^{\dagger}_{g,2}\hat{a}_{e,1})+\phi_{e,2}(x)\phi_{g,2}(x)(\hat{a}^{\dagger}_{e,2}\hat{a}_{g,2}+\hat{a}^{\dagger}_{g,2}\hat{a}_{e,2})+\phi_{e,2}(x)\phi_{g,3}(x)(\hat{a}^{\dagger}_{e,2}\hat{a}_{g,3}+\hat{a}^{\dagger}_{g,3}\hat{a}_{e,2})\right)\nonumber\\
&&\times(e^{ikx}\hat{b}_k +e^{-ikx}\hat{b}^\dagger_k )\nonumber\\
&&=\frac{2\pi a_{SB}}{\mu}\sqrt{\frac{\rho_B}{2m_B cV}}\sum_{k}\sqrt{k}\left((\hat{a}^{\dagger}_{e,1}\hat{a}_{g,1}+\hat{a}^{\dagger}_{g,1}\hat{a}_{e,1})\left(\hat{b}_{k}\int dx\;e^{ikx}\phi_{e,1}(x)\phi_{g,1}(x)\right.\right.\nonumber\\
&&\left.+\hat{b}^{\dagger}_{k}\int dx\;e^{-ikx}\phi_{e,1}(x)\phi_{g,1}(x)\right)+(\hat{a}^{\dagger}_{e,1}\hat{a}_{g,2}+\hat{a}^{\dagger}_{g,2}\hat{a}_{e,1})\left(\hat{b}_{k}\int dx\;e^{ikx}\phi_{e,1}(x)\phi_{g,2}(x)\right.\nonumber\\
&&\left.+\hat{b}^{\dagger}_{k}\int dx\;e^{-ikx}\phi_{e,1}(x)\phi_{g,2}(x)\right)+(\hat{a}^{\dagger}_{e,2}\hat{a}_{g,2}+\hat{a}^{\dagger}_{g,2}\hat{a}_{e,2})\left(\hat{b}_{k}\int dx\;e^{ikx}\phi_{e,2}(x)\phi_{g,2}(x)\right.\nonumber\\
&&\left.+\hat{b}^{\dagger}_{k}\int dx\;e^{-ikx}\phi_{e,2}(x)\phi_{g,2}(x)\right)+(\hat{a}^{\dagger}_{e,2}\hat{a}_{g,3}+\hat{a}^{\dagger}_{g,3}\hat{a}_{e,2})\left(\hat{b}_{k}\int dx\;e^{ikx}\phi_{e,2}(x)\phi_{g,3}(x)\right.\nonumber\\
&&\left.\left.+\hat{b}^{\dagger}_{k}\int dx\;e^{-ikx}\phi_{e,2}(x)\phi_{g,3}(x)\right)\right)\nonumber\\
\label{interactham1}
\end{eqnarray}

In this expression for the interaction Hamiltonian, we see that there are a number of overlap integrals of the form $\int dx\;e^{\pm i\mathbf{k}\cdot\mathbf{r}}\phi_{n,j}(x)\phi_{n',j'}(x)$. Substituting the expressions for the ground state and excited state eigenstates given by Eqs. \ref{grndstateeigen} and \ref{exctdstateeigen} into the overlap integrals, we find that

\begin{eqnarray}
&&\int dx\;e^{\pm ikx}\phi_{e,1}(x)\phi_{g,1}(x)=2\sigma_{g}^{1/2}\sigma_{e}^{-3/4}\int dx\;e^{\pm ikx}x\exp\left(-\frac{x^2}{2\sigma_e^2}\right)\delta(x+x_0)\nonumber\\
&&=-\frac{2}{\sigma^{1/4}_e}\sqrt{\frac{\sigma_g}{\sigma_e}}e^{\mp ikx_0}x_0\exp\left(-\frac{x_0^2}{2\sigma_e^2}\right)\nonumber\\
&&\int dx\;e^{\pm ikx}\phi_{e,1}(x)\phi_{g,2}(x)=2\sigma_{g}^{1/2}\sigma_{e}^{-3/4}\int dx\;e^{\pm ikx}x\exp\left(-\frac{x^2}{2\sigma_e^2}\right)\delta(x-x_0)\nonumber\\
&&=\frac{2}{\sigma^{1/4}_e}\sqrt{\frac{\sigma_g}{\sigma_e}}e^{\pm ikx_0}x_0\exp\left(-\frac{x_0^2}{2\sigma_e^2}\right)\nonumber\\
&&\int dx\;e^{\pm ikx}\phi_{e,2}(x)\phi_{g,2}(x)=2\sigma_{g}^{1/2}\sigma_{e}^{-3/4}\int dx\;e^{\pm ikx}(x-2x_0 )\exp\left(-\frac{(x-2x_0 )^2}{2\sigma_e^2}\right)\delta(x-x_0)\nonumber\\
&&=-\frac{2}{\sigma^{1/4}_e}\sqrt{\frac{\sigma_g}{\sigma_e}}e^{\pm 2ikx_0}x_0\exp\left(-\frac{x_0^2}{2\sigma_e^2}\right)\nonumber\\
&&\int dx\;e^{\pm ikx}\phi_{e,2}(x)\phi_{g,3}(x)=2\sigma_{g}^{1/2}\sigma_{e}^{-3/4}\int dx\;e^{\pm ikx}(x-2x_0 )\exp\left(-\frac{(x-2x_0 )^2}{2\sigma_e^2}\right)\delta(x-3x_0)\nonumber\\
&&=\frac{2}{\sigma^{1/4}_e}\sqrt{\frac{\sigma_g}{\sigma_e}}e^{\pm 3ikx_0}x_0\exp\left(-\frac{x_0^2}{2\sigma_e^2}\right)\nonumber\\
\label{overlapints} 
\end{eqnarray}
Now before we use these evaluated overlap integrals to simplify the interaction Hamiltonian between the trapped ultracold atom system and the background BEC, we first use the approximation that $kx_0 <<1$ for all $k$. As per Ref. \cite{caballar}, this approximation is made to ensure that we have inter-harmonic trap coherence, i. e. that each harmonic trap in the array is identical to each other. This is consistent with our assumption that the Bogoliubov excitations emitted by the ultracold atoms as they return to the ground state energy in the trap array are phononic, or have energies that vary linearly with $k$, i. e. $E_k=ck$, since under this approximation, the wave number $k$ of the excitations must be very small, while the inter-harmonic oscillator lengths $x_0$ must be very large, or vice versa. The former condition ensures that the excitation spectrum of the BEC, given by
\begin{equation}
E_k=ck\sqrt{1+\frac{k^2}{2m_b c}}
\label{excitationspec}
\end{equation}
can, via its Fourier expansion, be approximated as $E_k\approx ck$. At the same time, to avoid tunneling between traps, we must make $x_0$ much larger than the de Broglie wavelength of the ultracold atoms. Therefore, given these orders of magnitude for both $k$ and $x_0$, we then obtain the approximation mentioned earlier, $kx_0<<1$. Under this approximation, $e^{\pm ikx_0}\approx 1$, so that these factors which are present in the overlap integrals as given in Eq. \ref{overlapints} can be neglected. Thus, following this condition, we can then substitute the evaluated overlap integrals, sans the factors $e^{\pm ikx_0}$, into Eq \ref{interactham1} to obtain the following form of the interaction Hamiltonian: 

\begin{eqnarray}
&&\hat{H}_{SB}=\frac{4\pi a_{SB}}{\mu \sigma_e^{1/4}}\sqrt{\frac{\rho_B}{2m_B cL}\frac{\sigma_g}{\sigma_e}}x_0 \exp\left(-\frac{x_0^2}{2\sigma_e^2}\right)\sum_{k}\sqrt{k}\left(-(\hat{a}^{\dagger}_{e,1}\hat{a}_{g,1}+\hat{a}^{\dagger}_{g,1}\hat{a}_{e,1})\right.\nonumber\\
&&+(\hat{a}^{\dagger}_{e,1}\hat{a}_{g,2}+\hat{a}^{\dagger}_{g,2}\hat{a}_{e,1})-(\hat{a}^{\dagger}_{e,2}\hat{a}_{g,2}+\hat{a}^{\dagger}_{g,2}\hat{a}_{e,2})\nonumber\\
&&\left.+(\hat{a}^{\dagger}_{e,2}\hat{a}_{g,3}+\hat{a}^{\dagger}_{g,3}\hat{a}_{e,2})\right)(\hat{b}_{k}+\hat{b}^{\dagger}_{k})\nonumber\\
\label{interactham2}
\end{eqnarray}

We then evolve this Hamiltonian over time, making use of the Baker - Campbell - Hausdorff (BCH) identity as well as the commutation relations between the system and the BEC creation and annihilation operators in doing so. Explicitly, using the BCH identity, a time - evolved quantum operator $\hat{A}$ in a Hilbert space is given as

\begin{equation}
\hat{A}(t)=\exp\left(\frac{i}{\hbar}t\hat{H}\right)\hat{A}\exp\left(-\frac{i}{\hbar}t\hat{H}\right)=\hat{A}+\frac{i}{\hbar}t\left[\hat{H},\hat{A}\right]+\frac{1}{2!}\left(\frac{i}{\hbar}t\right)^2 \left[\hat{H},\left[\hat{H},\hat{A}\right]\right]+...
\label{bch}
\end{equation} 

On the other hand, the creation and annihilation operators for the system and the background BEC obey the following commutation relations:

\begin{equation}
\left[\hat{a}_{n,j},\hat{a}^{\dagger}_{n',j'}\right]=\delta_{n,n'}\delta_{j,j'},\;\left[\hat{b}_{k},\hat{b}^{\dagger}_{k'}\right]=\delta_{k,k'}
\label{creatannihilops}
\end{equation}

Thus, substituting the interaction Hamiltonian given by Eq. \ref{interactham2} into Eq. \ref{bch}, with the Hamiltonian term in the latter being the sum of the system and background BEC Hamiltonians, $\hat{H}=\hat{H}_S +\hat{H}_B$, with $\hat{H}_S$ and $\hat{H}_B$ given by Eqs. \ref{sysham} and \ref{becham}, respectively, and making use of the commutation relations given by Eq. \ref{creatannihilops} to evaluate the resulting commutators for the system and bath creation and annihilation operators, the time - evolved interaction Hamiltonian will then have the following form:

\begin{eqnarray}
&&\hat{H}_{SB}(t)=\frac{4\pi a_{SB}}{\mu \sigma_e^{1/4}}\sqrt{\frac{\rho_B}{2m_B cL}\frac{\sigma_g}{\sigma_e}}x_0 \exp\left(-\frac{x_0^2}{2\sigma_e^2}\right)\sum_{k}\sqrt{k}\nonumber\\
&&\times\left(-(e^{\frac{it}{\hbar}(\varepsilon_{e,1}-\varepsilon_{g,1})}\hat{a}^{\dagger}_{e,1}\hat{a}_{g,1}+e^{-\frac{it}{\hbar}(\varepsilon_{e,1}-\varepsilon_{g,1})}\hat{a}^{\dagger}_{g,1}\hat{a}_{e,1})\right.\nonumber\\
&&+(e^{\frac{it}{\hbar}(\varepsilon_{e,1}-\varepsilon_{g,2})}\hat{a}^{\dagger}_{e,1}\hat{a}_{g,2}+e^{-\frac{it}{\hbar}(\varepsilon_{e,1}-\varepsilon_{g,2})}\hat{a}^{\dagger}_{g,2}\hat{a}_{e,1})-(e^{\frac{it}{\hbar}(\varepsilon_{e,2}-\varepsilon_{g,2})}\hat{a}^{\dagger}_{e,2}\hat{a}_{g,2}+e^{-\frac{it}{\hbar}(\varepsilon_{e,2}-\varepsilon_{g,2})}\hat{a}^{\dagger}_{g,2}\hat{a}_{e,2})\nonumber\\
&&\left.+(e^{\frac{it}{\hbar}(\varepsilon_{e,2}-\varepsilon_{g,3})}\hat{a}^{\dagger}_{e,2}\hat{a}_{g,3}+e^{-\frac{it}{\hbar}(\varepsilon_{e,2}-\varepsilon_{g,3})}\hat{a}^{\dagger}_{g,3}\hat{a}_{e,2})\right)(e^{-\frac{it}{\hbar}E_k}\hat{b}_{k}+e^{\frac{it}{\hbar}E_k}\hat{b}^{\dagger}_{k})\nonumber\\
\label{interacthamevo}
\end{eqnarray}

\section{Derivation of the Master Equation for the Dissipative Dynamics of the System}

Having derived the interaction Hamiltonian describing the interaction between the trapped ultracold atom system and the background BEC, we now proceed to derive the master equation that describes the dissipative dynamics of the system. The general form of the master equation under the Born - Markov approximation is given as (see, for example, Ref. \cite{BreuerPetruccione})

\begin{equation}
\frac{d}{dt}\hat{\rho}_S = -\int_{0}^{\infty}dt'\;\mathrm{Tr}_{B}\left[\hat{H}_{SB}(t),\left[\hat{H}_{SB}(t-t'),\hat{\rho}_{S}(t)\otimes\hat{\rho}_B\right]\right]
\label{masteqgenform}
\end{equation}

Expanding the commutation relations in the integrand, this master equation can be written as

\begin{eqnarray}
&&\frac{d}{dt}\hat{\rho}_S = -\int_{0}^{\infty}dt'\;\mathrm{Tr}_{B}\left(\hat{H}_{SB}(t)\hat{H}_{SB}(t-t')\hat{\rho}_{S}(t)\otimes\hat{\rho}_B\right)\nonumber\\
&&+\int_{0}^{\infty}dt'\;\mathrm{Tr}_{B}\left(\hat{H}_{SB}(t)\hat{\rho}_{S}(t)\otimes\hat{\rho}_B\hat{H}_{SB}(t-t')\right)+\int_{0}^{\infty}dt'\;\mathrm{Tr}_{B}\left(\hat{H}_{SB}(t-t')\hat{\rho}_{S}(t)\otimes\hat{\rho}_B\hat{H}_{SB}(t)\right)\nonumber\\
&&-\int_{0}^{\infty}dt'\;\mathrm{Tr}_{B}\left(\hat{\rho}_{S}(t)\otimes\hat{\rho}_B\hat{H}_{SB}(t-t')\hat{H}_{SB}(t)\right)\nonumber\\
\label{masteqgenformexp}
\end{eqnarray}

In evaluating the commutators and the trace with respect to the background BEC variables in the integrand, we make use of the following identities, following Ref. \cite{daley}:

\begin{equation}
\mathrm{Tr}_{B}\left(\hat{b}_k \hat{b}^{\dagger}_{k'}\hat{\rho}_B\right)=\delta_{k,k'},\;\mathrm{Tr}_{B}\left(\hat{b}^{\dagger}_k \hat{b}_{k'}\hat{\rho}_B\right)=0,\;\mathrm{Tr}_{B}\left(\hat{b}_k \hat{b}_{k'}\hat{\rho}_B\right)=\mathrm{Tr}_{B}\left(\hat{b}^{\dagger}_k \hat{b}^{\dagger}_{k'}\hat{\rho}_B\right)=0
\label{bectraceidents}
\end{equation}
These identities are implemented to ensure that no excitations are emitted from the BEC to the ultracold atom gas to couple the ground states to the excited states (with excitation emission by the BEC corresponding to the second identity), while at the same time ensuring that the ultracold atoms in the excited energy state are able to decay back to one of the three available ground states in the trap via emission of excitations into the BEC (with excitation absorption by the BEC corresponding to the first identity). At the same time, we can make use of the cyclic nature of the trace, i. e. $\mathrm{Tr}_{B}\left(\hat{b}_k \hat{b}^{\dagger}_{k'}\hat{\rho}_B\right)=\mathrm{Tr}_{B}\left(\hat{b}^{\dagger}_{k'}\hat{\rho}_B\hat{b}_k\right)=\mathrm{Tr}_{B}\left(\hat{\rho}_B\hat{b}_k\hat{b}^{\dagger}_{k'}\right)=\delta_{k,k'}$. Thus, evaluating the commutator and the trace over the background BEC variables, and recalling that $\varepsilon_{g,1}=\varepsilon_{g,2}=\varepsilon_{g,3}=\varepsilon_g$ and $\varepsilon_{e,1}=\varepsilon_{e,2}=\varepsilon_e$ we obtain the following terms in the integrand:

\begin{eqnarray}
&&\mathrm{Tr}_{B}\left(\hat{H}_{SB}(t)\hat{H}_{SB}(t-t')\hat{\rho}_{S}(t)\otimes\hat{\rho}_B\right)=\frac{8\pi^2 a^2_{SB}}{\mu^2 \sigma_e^{1/2}}\frac{\rho_B}{m_B cL}\frac{\sigma_g}{\sigma_e}x^2_0 \exp\left(-\frac{x_0^2}{\sigma_e^2}\right)\sum_{k}ke^{\frac{it'}{\hbar}(\varepsilon_{e}-\varepsilon_{g}-ck)}\nonumber\\
&&\times\left((\hat{a}^{\dagger}_{e,1}\hat{a}_{g,1}+e^{-\frac{2it}{\hbar}(\varepsilon_{e}-\varepsilon_{g})}\hat{a}^{\dagger}_{g,1}\hat{a}_{e,1})-(\hat{a}^{\dagger}_{e,1}\hat{a}_{g,2}+e^{-\frac{2it}{\hbar}(\varepsilon_{e}-\varepsilon_{g})}\hat{a}^{\dagger}_{g,2}\hat{a}_{e,1})\right.\nonumber\\
&&\left.+(\hat{a}^{\dagger}_{e,2}\hat{a}_{g,2}+e^{-\frac{2it}{\hbar}(\varepsilon_{e}-\varepsilon_{g})}\hat{a}^{\dagger}_{g,2}\hat{a}_{e,2})-(\hat{a}^{\dagger}_{e,2}\hat{a}_{g,3}+e^{-\frac{2it}{\hbar}(\varepsilon_{e}-\varepsilon_{g})}\hat{a}^{\dagger}_{g,3}\hat{a}_{e,2})\right)\nonumber\\
&&\times\left((e^{\frac{2i(t-t')}{\hbar}(\varepsilon_{e}-\varepsilon_{g})}\hat{a}^{\dagger}_{e,1}\hat{a}_{g,1}+\hat{a}^{\dagger}_{g,1}\hat{a}_{e,1})-(e^{\frac{2i(t-t')}{\hbar}(\varepsilon_{e}-\varepsilon_{g})}\hat{a}^{\dagger}_{e,1}\hat{a}_{g,2}+\hat{a}^{\dagger}_{g,2}\hat{a}_{e,1})\right.\nonumber\\
&&\left.+(e^{\frac{2i(t-t')}{\hbar}(\varepsilon_{e}-\varepsilon_{g})}\hat{a}^{\dagger}_{e,2}\hat{a}_{g,2}+\hat{a}^{\dagger}_{g,2}\hat{a}_{e,2})-(e^{\frac{2i(t-t')}{\hbar}(\varepsilon_{e}-\varepsilon_{g})}\hat{a}^{\dagger}_{e,2}\hat{a}_{g,3}+\hat{a}^{\dagger}_{g,3}\hat{a}_{e,2})\right)\hat{\rho}_{S}(t)\nonumber\\
\label{integrandterm1}
\end{eqnarray}

\begin{eqnarray}
&&\mathrm{Tr}_{B}\left(\hat{H}_{SB}(t)\hat{\rho}_{S}(t)\otimes\hat{\rho}_{B}\hat{H}_{SB}(t-t')\right)=\frac{8\pi^2 a^2_{SB}}{\mu^2 \sigma_e^{1/2}}\frac{\rho_B}{m_B cL}\frac{\sigma_g}{\sigma_e}x^2_0 \exp\left(-\frac{x_0^2}{\sigma_e^2}\right)\sum_{k}ke^{\frac{it'}{\hbar}(\varepsilon_{e}-\varepsilon_{g}+ck)}\nonumber\\
&&\times\left((\hat{a}^{\dagger}_{e,1}\hat{a}_{g,1}+e^{-\frac{2it}{\hbar}(\varepsilon_{e}-\varepsilon_{g})}\hat{a}^{\dagger}_{g,1}\hat{a}_{e,1})-(\hat{a}^{\dagger}_{e,1}\hat{a}_{g,2}+e^{-\frac{2it}{\hbar}(\varepsilon_{e}-\varepsilon_{g})}\hat{a}^{\dagger}_{g,2}\hat{a}_{e,1})\right.\nonumber\\
&&\left.+(\hat{a}^{\dagger}_{e,2}\hat{a}_{g,2}+e^{-\frac{2it}{\hbar}(\varepsilon_{e}-\varepsilon_{g})}\hat{a}^{\dagger}_{g,2}\hat{a}_{e,2})-(\hat{a}^{\dagger}_{e,2}\hat{a}_{g,3}+e^{-\frac{2it}{\hbar}(\varepsilon_{e}-\varepsilon_{g})}\hat{a}^{\dagger}_{g,3}\hat{a}_{e,2})\right)\hat{\rho}_{S}(t)\nonumber\\
&&\times\left((e^{\frac{2i(t-t')}{\hbar}(\varepsilon_{e}-\varepsilon_{g})}\hat{a}^{\dagger}_{e,1}\hat{a}_{g,1}+\hat{a}^{\dagger}_{g,1}\hat{a}_{e,1})-(e^{\frac{2i(t-t')}{\hbar}(\varepsilon_{e}-\varepsilon_{g})}\hat{a}^{\dagger}_{e,1}\hat{a}_{g,2}+\hat{a}^{\dagger}_{g,2}\hat{a}_{e,1})\right.\nonumber\\
&&\left.+(e^{\frac{2i(t-t')}{\hbar}(\varepsilon_{e}-\varepsilon_{g})}\hat{a}^{\dagger}_{e,2}\hat{a}_{g,2}+\hat{a}^{\dagger}_{g,2}\hat{a}_{e,2})-(e^{\frac{2i(t-t')}{\hbar}(\varepsilon_{e}-\varepsilon_{g})}\hat{a}^{\dagger}_{e,2}\hat{a}_{g,3}+\hat{a}^{\dagger}_{g,3}\hat{a}_{e,2})\right)\nonumber\\
\label{integrandterm2}
\end{eqnarray}

\begin{eqnarray}
&&\mathrm{Tr}_{B}\left(\hat{H}_{SB}(t-t')\hat{\rho}_{S}(t)\otimes\hat{\rho}_{B}\hat{H}_{SB}(t)\right)=\frac{8\pi^2 a^2_{SB}}{\mu^2 \sigma_e^{1/2}}\frac{\rho_B}{m_B cL}\frac{\sigma_g}{\sigma_e}x^2_0 \exp\left(-\frac{x_0^2}{\sigma_e^2}\right)\sum_{k}ke^{-\frac{it'}{\hbar}(\epsilon_e-\epsilon_g+ck)}\nonumber\\
&&\times\left((\hat{a}^{\dagger}_{e,1}\hat{a}_{g,1}+e^{-\frac{2i(t-t')}{\hbar}(\varepsilon_{e}-\varepsilon_{g})}\hat{a}^{\dagger}_{g,1}\hat{a}_{e,1})-(\hat{a}^{\dagger}_{e,1}\hat{a}_{g,2}+e^{-\frac{2i(t-t')}{\hbar}(\varepsilon_{e}-\varepsilon_{g})}\hat{a}^{\dagger}_{g,2}\hat{a}_{e,1})\right.\nonumber\\
&&\left.+(\hat{a}^{\dagger}_{e,2}\hat{a}_{g,2}+e^{-\frac{2i(t-t')}{\hbar}(\varepsilon_{e}-\varepsilon_{g})}\hat{a}^{\dagger}_{g,2}\hat{a}_{e,2})-(\hat{a}^{\dagger}_{e,2}\hat{a}_{g,3}+e^{-\frac{2i(t-t')}{\hbar}(\varepsilon_{e}-\varepsilon_{g})}\hat{a}^{\dagger}_{g,3}\hat{a}_{e,2})\right)\hat{\rho}_{S}(t)\nonumber\\
&&\times\left((e^{\frac{2it}{\hbar}(\varepsilon_{e}-\varepsilon_{g})}\hat{a}^{\dagger}_{e,1}\hat{a}_{g,1}+\hat{a}^{\dagger}_{g,1}\hat{a}_{e,1})-(e^{\frac{2it}{\hbar}(\varepsilon_{e}-\varepsilon_{g})}\hat{a}^{\dagger}_{e,1}\hat{a}_{g,2}+\hat{a}^{\dagger}_{g,2}\hat{a}_{e,1})\right.\nonumber\\
&&\left.+(e^{\frac{2it}{\hbar}(\varepsilon_{e}-\varepsilon_{g})}\hat{a}^{\dagger}_{e,2}\hat{a}_{g,2}+\hat{a}^{\dagger}_{g,2}\hat{a}_{e,2})-(e^{\frac{2it}{\hbar}(\varepsilon_{e}-\varepsilon_{g})}\hat{a}^{\dagger}_{e,2}\hat{a}_{g,3}+\hat{a}^{\dagger}_{g,3}\hat{a}_{e,2})\right)\nonumber\\
\label{integrandterm3}
\end{eqnarray}

\begin{eqnarray}
&&\mathrm{Tr}_{B}\left(\hat{\rho}_{S}(t)\otimes\hat{\rho}_{B}\hat{H}_{SB}(t-t')\hat{H}_{SB}(t)\right)=\frac{8\pi^2 a^2_{SB}}{\mu^2 \sigma_e^{1/2}}\frac{\rho_B}{m_B cL}\frac{\sigma_g}{\sigma_e}x^2_0 \exp\left(-\frac{x_0^2}{\sigma_e^2}\right)\sum_{k}ke^{-\frac{it'}{\hbar}(\varepsilon_e - \varepsilon_g -ck)}\nonumber\\
&&\times\hat{\rho}_{S}(t)\left((\hat{a}^{\dagger}_{e,1}\hat{a}_{g,1}+e^{-\frac{2i(t-t')}{\hbar}(\varepsilon_{e}-\varepsilon_{g})}\hat{a}^{\dagger}_{g,1}\hat{a}_{e,1})-(\hat{a}^{\dagger}_{e,1}\hat{a}_{g,2}+e^{-\frac{2i(t-t')}{\hbar}(\varepsilon_{e}-\varepsilon_{g})}\hat{a}^{\dagger}_{g,2}\hat{a}_{e,1})\right.\nonumber\\
&&\left.+(\hat{a}^{\dagger}_{e,2}\hat{a}_{g,2}+e^{-\frac{2i(t-t')}{\hbar}(\varepsilon_{e}-\varepsilon_{g})}\hat{a}^{\dagger}_{g,2}\hat{a}_{e,2})-(\hat{a}^{\dagger}_{e,2}\hat{a}_{g,3}+e^{-\frac{2i(t-t')}{\hbar}(\varepsilon_{e}-\varepsilon_{g})}\hat{a}^{\dagger}_{g,3}\hat{a}_{e,2})\right)\nonumber\\
&&\times\left((e^{\frac{2it}{\hbar}(\varepsilon_{e}-\varepsilon_{g})}\hat{a}^{\dagger}_{e,1}\hat{a}_{g,1}+\hat{a}^{\dagger}_{g,1}\hat{a}_{e,1})-(e^{\frac{2it}{\hbar}(\varepsilon_{e}-\varepsilon_{g})}\hat{a}^{\dagger}_{e,1}\hat{a}_{g,2}+\hat{a}^{\dagger}_{g,2}\hat{a}_{e,1})\right.\nonumber\\
&&\left.+(e^{\frac{2it}{\hbar}(\varepsilon_{e}-\varepsilon_{g})}\hat{a}^{\dagger}_{e,2}\hat{a}_{g,2}+\hat{a}^{\dagger}_{g,2}\hat{a}_{e,2})-(e^{\frac{2it}{\hbar}(\varepsilon_{e}-\varepsilon_{g})}\hat{a}^{\dagger}_{e,2}\hat{a}_{g,3}+\hat{a}^{\dagger}_{g,3}\hat{a}_{e,2})\right)\nonumber\\
\label{integrandterm4}
\end{eqnarray}

Having calculated the explicit form of the integrands in Eq. \ref{masteqgenformexp}, we now proceed to simplify these terms. To do so, we make use of adiabatic elimination to express the creation and annihilation operators for the excited states of the system as linear combinations of the creation and annihilation operators of its ground states, as follows:

\begin{equation}
\hat{a}_{e,1}=\frac{\Omega_1}{\sqrt{2}\Delta_1}(\hat{a}_{g,1}+\hat{a}_{g,2}),\;\hat{a}_{e,2}=\frac{\Omega_2}{\sqrt{2}\Delta_2}(\hat{a}_{g,2}+\hat{a}_{g,3})
\label{adiabaticelimops}
\end{equation}

Substituting these operators into Eqs. \ref{integrandterm1} to \ref{integrandterm4} and simplifying the resulting expressions, we then obtain

\begin{eqnarray}
&&\mathrm{Tr}_{B}\left(\hat{H}_{SB}(t)\hat{H}_{SB}(t-t')\hat{\rho}_{S}(t)\otimes\hat{\rho}_B\right)=\frac{4\pi^2 a^2_{SB}}{\mu^2 \sqrt{\sigma_e}}\frac{\rho_B}{m_B cL}\frac{\sigma_g}{\sigma_e}\left(\frac{\Omega_1}{\Delta_1}x_0\right)^2 \exp\left(-\frac{x_0^2}{\sigma_e^2}\right)\nonumber\\
&&\times\sum_{k}ke^{\frac{it'}{\hbar}(\varepsilon_{e}-\varepsilon_{g}-ck)}\left(((\hat{a}^{\dagger}_{g,1}+\hat{a}^{\dagger}_{g,2})(\hat{a}_{g,1}-\hat{a}_{g,2})+e^{-\frac{2it}{\hbar}(\varepsilon_{e}-\varepsilon_{g})}(\hat{a}^{\dagger}_{g,1}-\hat{a}^{\dagger}_{g,2})(\hat{a}_{g,1}+\hat{a}_{g,2}))\right.\nonumber\\
&&\left.+\epsilon((\hat{a}^{\dagger}_{g,2}+\hat{a}^{\dagger}_{g,3})(\hat{a}_{g,2}-\hat{a}_{g,3})+e^{-\frac{2it}{\hbar}(\varepsilon_{e}-\varepsilon_{g})}(\hat{a}^{\dagger}_{g,2}-\hat{a}^{\dagger}_{g,3})(\hat{a}_{g,2}+\hat{a}_{g,3}))\right)\nonumber\\
&&\times\left((e^{\frac{2i(t-t')}{\hbar}(\varepsilon_{e}-\varepsilon_{g})}(\hat{a}^{\dagger}_{g,1}+\hat{a}^{\dagger}_{g,2})(\hat{a}_{g,1}-\hat{a}_{g,2})+(\hat{a}^{\dagger}_{g,1}-\hat{a}^{\dagger}_{g,2})(\hat{a}_{g,1}+\hat{a}_{g,2}))\right.\nonumber\\
&&\left.+\epsilon(e^{\frac{2i(t-t')}{\hbar}(\varepsilon_{e}-\varepsilon_{g})}(\hat{a}^{\dagger}_{g,2}+\hat{a}^{\dagger}_{g,3})(\hat{a}_{g,2}-\hat{a}_{g,3})+(\hat{a}^{\dagger}_{g,2}-\hat{a}^{\dagger}_{g,3})(\hat{a}_{g,2}+\hat{a}_{g,3})\right)\hat{\rho}_{S}(t)\nonumber\\
\label{integrandterm1a}
\end{eqnarray}

\begin{eqnarray}
&&\mathrm{Tr}_{B}\left(\hat{H}_{SB}(t)\hat{\rho}_{S}(t)\otimes\hat{\rho}_{B}\hat{H}_{SB}(t-t')\right)=\frac{4\pi^2 a^2_{SB}}{\mu^2 \sqrt{\sigma_e}}\frac{\rho_B}{m_B cL}\frac{\sigma_g}{\sigma_e}\left(\frac{\Omega_1}{\Delta_1}x_0\right)^2 \exp\left(-\frac{x_0^2}{\sigma_e^2}\right)\nonumber\\
&&\times\sum_{k}ke^{\frac{it'}{\hbar}(\varepsilon_{e}-\varepsilon_{g}+ck)}\left(((\hat{a}^{\dagger}_{g,1}+\hat{a}^{\dagger}_{g,2})(\hat{a}_{g,1}-\hat{a}_{g,2})+e^{-\frac{2it}{\hbar}(\varepsilon_{e}-\varepsilon_{g})}(\hat{a}^{\dagger}_{g,1}-\hat{a}^{\dagger}_{g,2})(\hat{a}_{g,1}+\hat{a}_{g,2}))\right.\nonumber\\
&&\left.+\epsilon((\hat{a}^{\dagger}_{g,2}+\hat{a}^{\dagger}_{g,3})(\hat{a}_{g,2}-\hat{a}_{g,3})+e^{-\frac{2it}{\hbar}(\varepsilon_{e}-\varepsilon_{g})}(\hat{a}^{\dagger}_{g,2}-\hat{a}^{\dagger}_{g,3})(\hat{a}_{g,2}+\hat{a}_{g,3}))\right)\hat{\rho}_{S}(t)\nonumber\\
&&\times\left((e^{\frac{2i(t-t')}{\hbar}(\varepsilon_{e}-\varepsilon_{g})}(\hat{a}^{\dagger}_{g,1}+\hat{a}^{\dagger}_{g,2})(\hat{a}_{g,1}-\hat{a}_{g,2})+(\hat{a}^{\dagger}_{g,1}-\hat{a}^{\dagger}_{g,2})(\hat{a}_{g,1}+\hat{a}_{g,2}))\right.\nonumber\\
&&\left.+\epsilon(e^{\frac{2i(t-t')}{\hbar}(\varepsilon_{e}-\varepsilon_{g})}(\hat{a}^{\dagger}_{g,2}+\hat{a}^{\dagger}_{g,3})(\hat{a}_{g,2}-\hat{a}_{g,3})+(\hat{a}^{\dagger}_{g,2}-\hat{a}^{\dagger}_{g,3})(\hat{a}_{g,2}+\hat{a}_{g,3}))\right)\nonumber\\
\label{integrandterm2a}
\end{eqnarray}

\begin{eqnarray}
&&\mathrm{Tr}_{B}\left(\hat{H}_{SB}(t-t')\hat{\rho}_{S}(t)\otimes\hat{\rho}_{B}\hat{H}_{SB}(t)\right)=\frac{4\pi^2 a^2_{SB}}{\mu^2 \sqrt{\sigma_e}}\frac{\rho_B}{m_B cL}\frac{\sigma_g}{\sigma_e}\left(\frac{\Omega_1}{\Delta_1}x_0\right)^2 \exp\left(-\frac{x_0^2}{\sigma_e^2}\right)\nonumber\\
&&\times\sum_{k}ke^{-\frac{it'}{\hbar}(\epsilon_e-\epsilon_g+ck)}\left(((\hat{a}^{\dagger}_{g,1}+\hat{a}^{\dagger}_{g,2})(\hat{a}_{g,1}-\hat{a}_{g,2})+e^{-\frac{2i(t-t')}{\hbar}(\varepsilon_{e}-\varepsilon_{g})}(\hat{a}^{\dagger}_{g,1}-\hat{a}^{\dagger}_{g,2})(\hat{a}_{g,1}+\hat{a}_{g,2}))\right.\nonumber\\
&&\left.+\epsilon((\hat{a}^{\dagger}_{g,2}+\hat{a}^{\dagger}_{g,3})(\hat{a}_{g,2}-\hat{a}_{g,3})+e^{-\frac{2i(t-t')}{\hbar}(\varepsilon_{e}-\varepsilon_{g})}(\hat{a}^{\dagger}_{g,2}-\hat{a}^{\dagger}_{g,3})(\hat{a}_{g,2}+\hat{a}_{g,3}))\right)\hat{\rho}_{S}(t)\nonumber\\
&&\times\left((e^{\frac{2it}{\hbar}(\varepsilon_{e}-\varepsilon_{g})}(\hat{a}^{\dagger}_{g,1}+\hat{a}^{\dagger}_{g,2})(\hat{a}_{g,1}-\hat{a}_{g,2})+(\hat{a}^{\dagger}_{g,1}-\hat{a}^{\dagger}_{g,2})(\hat{a}_{g,1}+\hat{a}_{g,2}))\right.\nonumber\\
&&\left.+\epsilon(e^{\frac{2it}{\hbar}(\varepsilon_{e}-\varepsilon_{g})}(\hat{a}^{\dagger}_{g,2}+\hat{a}^{\dagger}_{g,3})(\hat{a}_{g,2}-\hat{a}_{g,3})+(\hat{a}^{\dagger}_{g,2}-\hat{a}^{\dagger}_{g,3})(\hat{a}_{g,2}+\hat{a}_{g,3}))\right)\nonumber\\
\label{integrandterm3a}
\end{eqnarray}

\begin{eqnarray}
&&\mathrm{Tr}_{B}\left(\hat{\rho}_{S}(t)\otimes\hat{\rho}_{B}\hat{H}_{SB}(t-t')\hat{H}_{SB}(t)\right)=\frac{4\pi^2 a^2_{SB}}{\mu^2 \sqrt{\sigma_e}}\frac{\rho_B}{m_B cL}\frac{\sigma_g}{\sigma_e}\left(\frac{\Omega_1}{\Delta_1}x_0\right)^2 \exp\left(-\frac{x_0^2}{\sigma_e^2}\right)\nonumber\\
&&\times\sum_{k}ke^{-\frac{it'}{\hbar}(\varepsilon_e - \varepsilon_g -ck)}\hat{\rho}_{S}(t)\left(((\hat{a}^{\dagger}_{g,1}+\hat{a}^{\dagger}_{g,2})(\hat{a}_{g,1}-\hat{a}_{g,2})+e^{-\frac{2i(t-t')}{\hbar}(\varepsilon_{e}-\varepsilon_{g})}(\hat{a}^{\dagger}_{g,1}-\hat{a}_{g,2})(\hat{a}_{g,1}+\hat{a}_{g,2}))\right.\nonumber\\
&&\left.+\epsilon((\hat{a}^{\dagger}_{g,2}+\hat{a}^{\dagger}_{g,3})(\hat{a}_{g,2}-\hat{a}_{g,3})+e^{-\frac{2i(t-t')}{\hbar}(\varepsilon_{e}-\varepsilon_{g})}(\hat{a}^{\dagger}_{g,2}-\hat{a}^{\dagger}_{g,3})(\hat{a}_{g,2}+\hat{a}_{g,3}))\right)\nonumber\\
&&\times\left((e^{\frac{2it}{\hbar}(\varepsilon_{e}-\varepsilon_{g})}(\hat{a}^{\dagger}_{g,1}+\hat{a}^{\dagger}_{g,2})(\hat{a}_{g,1}-\hat{a}_{g,2})+(\hat{a}^{\dagger}_{g,1}-\hat{a}^{\dagger}_{g,2})(\hat{a}_{g,1}+\hat{a}_{g,2}))\right.\nonumber\\
&&\left.+\epsilon(e^{\frac{2it}{\hbar}(\varepsilon_{e}-\varepsilon_{g})}(\hat{a}^{\dagger}_{g,2}+\hat{a}^{\dagger}_{g,3})(\hat{a}_{g,2}-\hat{a}_{g,3})+(\hat{a}^{\dagger}_{g,2}-\hat{a}^{\dagger}_{g,3})(\hat{a}_{g,2}+\hat{a}_{g,3}))\right)\nonumber\\
\label{integrandterm4a}
\end{eqnarray}

To simplify these terms further, we now proceed to integrating them over time. At the same time, we replace the summation over $k$ with an integration over $k$, treating the summation variable as continuous to be able to do so. In doing so, we make use of the definition of the Dirac delta distribution in terms of oscillatory integrals, giving us the following integration identities:

\begin{eqnarray}
&&\sum_{k}\int_{0}^{+\infty}dt'e^{\pm\frac{i}{\hbar}(\varepsilon_e -\varepsilon_g -ck)t'}k'=\frac{\hbar}{\sqrt{2\pi}}\int_{-\infty}^{+\infty}dk'k'\delta(\varepsilon_e -\varepsilon_g -ck')=\frac{1}{\sqrt{2\pi}}\frac{\hbar}{c}(\varepsilon_e -\varepsilon_g)\nonumber\\
&&\sum_{k}\int_{0}^{+\infty}dt'e^{\pm\frac{i}{\hbar}(\varepsilon_e -\varepsilon_g+ck)t'}k'=\frac{\hbar}{\sqrt{2\pi}}\int_{-\infty}^{+\infty}dk'k'\delta(\varepsilon_e -\varepsilon_g+ck')=-\frac{1}{\sqrt{2\pi}}\frac{\hbar}{c}(\varepsilon_e -\varepsilon_g)\nonumber\\
\label{timeandkintegrals}
\end{eqnarray}

Thus, integrating the terms given by Eqs. \ref{integrandterm1a} to \ref{integrandterm4a} over time and over $k$, and making use of the integration identities given by Eq. \ref{timeandkintegrals} in doing so, we find that

\begin{eqnarray}
&&\int_{0}^{+\infty}dt'\;\mathrm{Tr}_{B}\left(\hat{H}_{SB}(t)\hat{H}_{SB}(t-t')\hat{\rho}_{S}(t)\otimes\hat{\rho}_B\right)\nonumber\\
&&=\frac{4\hbar\pi^2 a^2_{SB}}{\mu^2 \sqrt{2\pi\sigma_e}}\frac{\rho_B}{m_B c^2 L}\frac{\sigma_g}{\sigma_e}\left(\frac{\Omega_1}{\Delta_1}x_0\right)^2 (\varepsilon_e -\varepsilon_g) \exp\left(-\frac{x_0^2}{\sigma_e^2}\right)\nonumber\\
&&\times\left(((\hat{a}^{\dagger}_{g,1}+\hat{a}^{\dagger}_{g,2})(\hat{a}_{g,1}-\hat{a}_{g,2})+e^{-\frac{2it}{\hbar}(\varepsilon_{e}-\varepsilon_{g})}(\hat{a}^{\dagger}_{g,1}-\hat{a}^{\dagger}_{g,2})(\hat{a}_{g,1}+\hat{a}_{g,2}))\right.\nonumber\\
&&\left.+\epsilon((\hat{a}^{\dagger}_{g,2}+\hat{a}^{\dagger}_{g,3})(\hat{a}_{g,2}-\hat{a}_{g,3})+e^{-\frac{2it}{\hbar}(\varepsilon_{e}-\varepsilon_{g})}(\hat{a}^{\dagger}_{g,2}-\hat{a}^{\dagger}_{g,3})(\hat{a}_{g,2}+\hat{a}_{g,3}))\right)\nonumber\\
&&\times\left((-e^{\frac{2it}{\hbar}(\varepsilon_{e}-\varepsilon_{g})}(\hat{a}^{\dagger}_{g,1}+\hat{a}^{\dagger}_{g,2})(\hat{a}_{g,1}-\hat{a}_{g,2})+(\hat{a}^{\dagger}_{g,1}-\hat{a}^{\dagger}_{g,2})(\hat{a}_{g,1}+\hat{a}_{g,2}))\right.\nonumber\\
&&\left.+\epsilon(-e^{\frac{2it}{\hbar}(\varepsilon_{e}-\varepsilon_{g})}(\hat{a}^{\dagger}_{g,2}+\hat{a}^{\dagger}_{g,3})(\hat{a}_{g,2}-\hat{a}_{g,3})+(\hat{a}^{\dagger}_{g,2}-\hat{a}^{\dagger}_{g,3})(\hat{a}_{g,2}+\hat{a}_{g,3})\right)\hat{\rho}_{S}(t)\nonumber\\
\label{integrandterm1b}
\end{eqnarray}

\begin{eqnarray}
&&\int_{0}^{+\infty}dt'\;\mathrm{Tr}_{B}\left(\hat{H}_{SB}(t)\hat{\rho}_{S}(t)\otimes\hat{\rho}_{B}\hat{H}_{SB}(t-t')\right)\nonumber\\
&&=\frac{4\hbar\pi^2 a^2_{SB}}{\mu^2 \sqrt{2\pi\sigma_e}}\frac{\rho_B}{m_B c^2 L}\frac{\sigma_g}{\sigma_e}\left(\frac{\Omega_1}{\Delta_1}x_0\right)^2 (\varepsilon_e - \varepsilon_g)\exp\left(-\frac{x_0^2}{\sigma_e^2}\right)\nonumber\\
&&\left(((\hat{a}^{\dagger}_{g,1}+\hat{a}^{\dagger}_{g,2})(\hat{a}_{g,1}-\hat{a}_{g,2})+e^{-\frac{2it}{\hbar}(\varepsilon_{e}-\varepsilon_{g})}(\hat{a}^{\dagger}_{g,1}-\hat{a}^{\dagger}_{g,2})(\hat{a}_{g,1}+\hat{a}_{g,2}))\right.\nonumber\\
&&\left.+\epsilon((\hat{a}^{\dagger}_{g,2}+\hat{a}^{\dagger}_{g,3})(\hat{a}_{g,2}-\hat{a}_{g,3})+e^{-\frac{2it}{\hbar}(\varepsilon_{e}-\varepsilon_{g})}(\hat{a}^{\dagger}_{g,2}-\hat{a}^{\dagger}_{g,3})(\hat{a}_{g,2}+\hat{a}_{g,3}))\right)\hat{\rho}_{S}(t)\nonumber\\
&&\times\left((e^{\frac{2it}{\hbar}(\varepsilon_{e}-\varepsilon_{g})}(\hat{a}^{\dagger}_{g,1}+\hat{a}^{\dagger}_{g,2})(\hat{a}_{g,1}-\hat{a}_{g,2})-(\hat{a}^{\dagger}_{g,1}-\hat{a}^{\dagger}_{g,2})(\hat{a}_{g,1}+\hat{a}_{g,2}))\right.\nonumber\\
&&\left.+\epsilon(e^{\frac{2it}{\hbar}(\varepsilon_{e}-\varepsilon_{g})}(\hat{a}^{\dagger}_{g,2}+\hat{a}^{\dagger}_{g,3})(\hat{a}_{g,2}-\hat{a}_{g,3})-(\hat{a}^{\dagger}_{g,2}-\hat{a}^{\dagger}_{g,3})(\hat{a}_{g,2}+\hat{a}_{g,3}))\right)\nonumber\\
\label{integrandterm2b}
\end{eqnarray}

\begin{eqnarray}
&&\mathrm{Tr}_{B}\left(\hat{H}_{SB}(t-t')\hat{\rho}_{S}(t)\otimes\hat{\rho}_{B}\hat{H}_{SB}(t)\right)\nonumber\\
&&=\frac{4\hbar\pi^2 a^2_{SB}}{\mu^2 \sqrt{2\pi\sigma_e}}\frac{\rho_B}{m_B c^2 L}\frac{\sigma_g}{\sigma_e}\left(\frac{\Omega_1}{\Delta_1}x_0\right)^2 (\varepsilon_e - \varepsilon_g)\exp\left(-\frac{x_0^2}{\sigma_e^2}\right)\nonumber\\
&&\times\left((-(\hat{a}^{\dagger}_{g,1}+\hat{a}^{\dagger}_{g,2})(\hat{a}_{g,1}-\hat{a}_{g,2})+e^{-\frac{2it}{\hbar}(\varepsilon_{e}-\varepsilon_{g})}(\hat{a}^{\dagger}_{g,1}-\hat{a}^{\dagger}_{g,2})(\hat{a}_{g,1}+\hat{a}_{g,2}))\right.\nonumber\\
&&\left.+\epsilon(-(\hat{a}^{\dagger}_{g,2}+\hat{a}^{\dagger}_{g,3})(\hat{a}_{g,2}-\hat{a}_{g,3})+e^{-\frac{2it}{\hbar}(\varepsilon_{e}-\varepsilon_{g})}(\hat{a}^{\dagger}_{g,2}-\hat{a}^{\dagger}_{g,3})(\hat{a}_{g,2}+\hat{a}_{g,3}))\right)\hat{\rho}_{S}(t)\nonumber\\
&&\times\left((e^{\frac{2it}{\hbar}(\varepsilon_{e}-\varepsilon_{g})}(\hat{a}^{\dagger}_{g,1}+\hat{a}^{\dagger}_{g,2})(\hat{a}_{g,1}-\hat{a}_{g,2})+(\hat{a}^{\dagger}_{g,1}-\hat{a}^{\dagger}_{g,2})(\hat{a}_{g,1}+\hat{a}_{g,2}))\right.\nonumber\\
&&\left.+\epsilon(e^{\frac{2it}{\hbar}(\varepsilon_{e}-\varepsilon_{g})}(\hat{a}^{\dagger}_{g,2}+\hat{a}^{\dagger}_{g,3})(\hat{a}_{g,2}-\hat{a}_{g,3})+(\hat{a}^{\dagger}_{g,2}-\hat{a}^{\dagger}_{g,3})(\hat{a}_{g,2}+\hat{a}_{g,3}))\right)\nonumber\\
\label{integrandterm3b}
\end{eqnarray}

\begin{eqnarray}
&&\mathrm{Tr}_{B}\left(\hat{\rho}_{S}(t)\otimes\hat{\rho}_{B}\hat{H}_{SB}(t-t')\hat{H}_{SB}(t)\right)\nonumber\\
&&=\frac{4\hbar\pi^2 a^2_{SB}}{\mu^2 \sqrt{2\pi\sigma_e}}\frac{\rho_B}{m_B c^2 L}\frac{\sigma_g}{\sigma_e}\left(\frac{\Omega_1}{\Delta_1}x_0\right)^2 (\varepsilon_e - \varepsilon_g)\exp\left(-\frac{x_0^2}{\sigma_e^2}\right)\nonumber\\
&&\times\hat{\rho}_{S}(t)\left(((\hat{a}^{\dagger}_{g,1}+\hat{a}^{\dagger}_{g,2})(\hat{a}_{g,1}-\hat{a}_{g,2})-e^{-\frac{2it}{\hbar}(\varepsilon_{e}-\varepsilon_{g})}(\hat{a}^{\dagger}_{g,1}-\hat{a}_{g,2})(\hat{a}_{g,1}+\hat{a}_{g,2}))\right.\nonumber\\
&&\left.+\epsilon((\hat{a}^{\dagger}_{g,2}+\hat{a}^{\dagger}_{g,3})(\hat{a}_{g,2}-\hat{a}_{g,3})-e^{-\frac{2it}{\hbar}(\varepsilon_{e}-\varepsilon_{g})}(\hat{a}^{\dagger}_{g,2}-\hat{a}^{\dagger}_{g,3})(\hat{a}_{g,2}+\hat{a}_{g,3}))\right)\nonumber\\
&&\times\left((e^{\frac{2it}{\hbar}(\varepsilon_{e}-\varepsilon_{g})}(\hat{a}^{\dagger}_{g,1}+\hat{a}^{\dagger}_{g,2})(\hat{a}_{g,1}-\hat{a}_{g,2})+(\hat{a}^{\dagger}_{g,1}-\hat{a}^{\dagger}_{g,2})(\hat{a}_{g,1}+\hat{a}_{g,2}))\right.\nonumber\\
&&\left.+\epsilon(e^{\frac{2it}{\hbar}(\varepsilon_{e}-\varepsilon_{g})}(\hat{a}^{\dagger}_{g,2}+\hat{a}^{\dagger}_{g,3})(\hat{a}_{g,2}-\hat{a}_{g,3})+(\hat{a}^{\dagger}_{g,2}-\hat{a}^{\dagger}_{g,3})(\hat{a}_{g,2}+\hat{a}_{g,3}))\right)\nonumber\\
\label{integrandterm4b}
\end{eqnarray}

Now let us introduce the following jump operator to simplify the expressions from Eqs. \ref{integrandterm1b} to \ref{integrandterm4b}:

\begin{equation}
\hat{c}=(\hat{a}^{\dagger}_{g,1}+\hat{a}^{\dagger}_{g,2})(\hat{a}_{g,1}-\hat{a}_{g,2})+\epsilon(\hat{a}^{\dagger}_{g,2}+\hat{a}^{\dagger}_{g,3})(\hat{a}_{g,2}-\hat{a}_{g,3})
\label{jumpop}
\end{equation}
Also, we designate $A$ as a constant that denotes the strength of coupling between the system and the background BEC as follows:

\begin{equation}
A=\frac{4\hbar\pi^2 a^2_{SB}}{\mu^2 \sqrt{2\pi\sigma_e}}\frac{\rho_B}{m_B c^2 L}\frac{\sigma_g}{\sigma_e}\left(\frac{\Omega_1}{\Delta_1}x_0\right)^2 (\varepsilon_e - \varepsilon_g)\exp\left(-\frac{x_0^2}{\sigma_e^2}\right)
\label{couplingconst}
\end{equation}

Substituting this into Eqs. \ref{integrandterm1b} to \ref{integrandterm4b}, we then obtain

\begin{eqnarray}
&&\int_{0}^{+\infty}dt'\;\mathrm{Tr}_{B}\left(\hat{H}_{SB}(t)\hat{H}_{SB}(t-t')\hat{\rho}_{S}(t)\otimes\hat{\rho}_B\right)=A(\hat{c}+e^{-\frac{2it}{\hbar}(\varepsilon_{e}-\varepsilon_{g})}\hat{c}^{\dagger})(\hat{c}^{\dagger}-e^{\frac{2it}{\hbar}(\varepsilon_{e}-\varepsilon_{g})}\hat{c})\hat{\rho}_{S}(t)\nonumber\\
&&=A(\hat{c}\hat{c}^{\dagger}\hat{\rho}_{S}(t)-\hat{c}^{\dagger}\hat{c}\hat{\rho}_{S}(t)+e^{-\frac{2it}{\hbar}(\varepsilon_e -\varepsilon_g)}\hat{c}^{\dagger}\hat{c}^{\dagger}\hat{\rho}_{S}(t)-e^{\frac{2it}{\hbar}(\varepsilon_e -\varepsilon_g)}\hat{c}\hat{c}\hat{\rho}_{S}(t))\nonumber\\
\label{integrandterm1c}
\end{eqnarray}

\begin{eqnarray}
&&\int_{0}^{+\infty}dt'\;\mathrm{Tr}_{B}\left(\hat{H}_{SB}(t)\hat{\rho}_{S}(t)\otimes\hat{\rho}_{B}\hat{H}_{SB}(t-t')\right)=A(\hat{c}+e^{-\frac{2it}{\hbar}(\varepsilon_{e}-\varepsilon_{g})}\hat{c}^{\dagger})\hat{\rho}_{S}(t)(e^{\frac{2it}{\hbar}(\varepsilon_{e}-\varepsilon_{g})}\hat{c}-\hat{c}^{\dagger})\nonumber\\
&&=A(\hat{c}^{\dagger}\hat{\rho}_{S}(t)\hat{c}-\hat{c}\hat{\rho}_{S}(T)\hat{c}^{\dagger}-e^{-\frac{2it}{\hbar}(\varepsilon_e -\varepsilon_g)}\hat{c}^{\dagger}\hat{\rho}_{S}(t)\hat{c}^{\dagger}+e^{\frac{2it}{\hbar}(\varepsilon_e -\varepsilon_g)}\hat{c}\hat{\rho}_{S}(t)\hat{c})\nonumber\\
\label{integrandterm2c}
\end{eqnarray}

\begin{eqnarray}
&&\int_{0}^{+\infty}dt'\;\mathrm{Tr}_{B}\left(\hat{H}_{SB}(t-t')\hat{\rho}_{S}(t)\otimes\hat{\rho}_{B}\hat{H}_{SB}(t)\right)=A(-\hat{c}+e^{-\frac{2it}{\hbar}(\varepsilon_{e}-\varepsilon_{g})}\hat{c}^{\dagger})\hat{\rho}_{S}(t)(e^{\frac{2it}{\hbar}(\varepsilon_{e}-\varepsilon_{g})}\hat{c}+\hat{c}^{\dagger})\nonumber\\
&&=A(\hat{c}^{\dagger}\hat{\rho}_{S}(t)\hat{c}-\hat{c}\hat{\rho}_{S}(t)\hat{c}^{\dagger}+e^{-\frac{2it}{\hbar}(\varepsilon_e -\varepsilon_g)}\hat{c}^{\dagger}\hat{\rho}_{S}(t)\hat{c}^{\dagger}-e^{\frac{2it}{\hbar}(\varepsilon_e -\varepsilon_g)}\hat{c}\hat{\rho}_{S}(t)\hat{c})
\label{integrandterm3c}
\end{eqnarray}

\begin{eqnarray}
&&\int_{0}^{+\infty}dt'\;\mathrm{Tr}_{B}\left(\hat{\rho}_{S}(t)\otimes\hat{\rho}_{B}\hat{H}_{SB}(t-t')\hat{H}_{SB}(t)\right)=A\hat{\rho}_{S}(t)(\hat{c}-e^{-\frac{2it}{\hbar}(\varepsilon_{e}-\varepsilon_{g})}\hat{c}^{\dagger})(e^{\frac{2it}{\hbar}(\varepsilon_{e}-\varepsilon_{g})}\hat{c}+hat{c}^{\dagger})\nonumber\\
&&=A\hat{\rho}_{S}(t)(\hat{c}\hat{c}^{\dagger}-\hat{c}^{\dagger}\hat{c}+e^{\frac{2it}{\hbar}(\varepsilon_e -\varepsilon_g)}\hat{c}\hat{c}-e^{-\frac{2it}{\hbar}(\varepsilon_e -\varepsilon_g)}\hat{c}^{\dagger}\hat{c}^{\dagger})\nonumber\\
\label{integrandterm4c}
\end{eqnarray}

Finally, substituting Eqs. \ref{integrandterm1c} to \ref{integrandterm4c} into Eq. \ref{masteqgenformexp} and evaluating the resulting expression, we obtain the following form of the master equation governing the dissipative dynamics due to the interaction between the trapped ultracold atom system and the background BEC:

\begin{eqnarray}
&&\frac{d}{dt}\hat{\rho}_{S}(t)=A\left((2\hat{c}^{\dagger}\hat{\rho}_{S}(t)\hat{c}-\left\{\hat{c}\hat{c}^{\dagger},\hat{\rho}_{S}(t)\right\})-(2\hat{c}\hat{\rho}_{S}(t)\hat{c}^{\dagger}-\left\{\hat{c}^{\dagger}\hat{c},\hat{\rho}_{S}(t)\right\})\right.\nonumber\\
&&\left.+e^{\frac{2it}{\hbar}(\varepsilon_e -\varepsilon_g)}\left[\hat{c}\hat{c},\hat{\rho}_{S}(t)\right]-e^{-\frac{2it}{\hbar}(\varepsilon_e -\varepsilon_g)}\left[\hat{c}^{\dagger}\hat{c}^{\dagger},\hat{\rho}_{S}(t)\right]\right)\nonumber\\
\label{masteqsysbec}
\end{eqnarray}

\section{Time Evolution of the Trapped Ultracold Atom System Via the Master Equation}

Having derived the master equation that describes the driven - dissipative dynamics of the trapped ultracold atom system as it interacts with the background BEC, we now proceed to use this equation to evolve the system over time, and evaluate some physical properties that are of interest for this system. Before we do so, however, let us make some notes regarding the equation itself, in particular the coupling constant $A$, given by Eq. \ref{couplingconst}. First, the constant $A$ is known as the coupling constant because its magnitude specifies how strongly the trapped ultracold atom system interacts with the background BEC. The greater the magnitude of $A$, the stronger the interaction between the system and the BEC. At the same time, the explicit form of the coupling constant shows that the coupling strength between the trapped ultracold atom system and the background BEC can be adjusted by adjusting one or more physical parameters governing this coupling, such as the trap widths at the ground and excited states, $\sigma_g$ and $\sigma_e$, as well as the scattering length $a_SB$. However, the parameter that is easiest to access from an experimental standpoint would be the Rabi frequency $\Omega_1$ of the Raman laser coupling the ground states located in the first and third harmonic traps to the excited state in the second harmonic trap, which can be done by replacing the Raman laser coupling these energy levels with another of a different frequency. We note that changing this Rabi frequency $\Omega_1$ also changes the detuning $\Delta_1$ of the Raman laser. 

This ability to vary the coupling strength between the trapped ultracold atom system and the background BEC is crucial to the manner by which the coupled trapped ultracold atom - background BEC system will be evolved over time. This is because, following Refs. \cite{caballar} and \cite{caballar2}, over the course of the time evolution of the trapped ultracold atoms, we will be carrying out a stroboscopic coupling of the trapped ultracold atoms and the background BEC, wherein we vary the strength of the coupling between the trapped atoms and the background BEC at particular instants of time. In particular, we couple the ground states and the excited states of the ultracold atoms in their respective trap locations using the Raman lasers while they interact with the background BEC for only a finite amount of time before we turn off the Raman lasers and replace one of them, specifically the first Raman laser which couples the ground states in the first and third harmonic traps to the excited state of the second excited trap, with another Raman laser with a different Rabi frequency, afterwhich we turn both lasers on again to couple the energy levels and repeat the process over a number of intervals of time. This stroboscopic coupling results in the trapped ultracold atoms evolving via a combination of driven and dissipative dynamics towards a steady state, as shown in the figures below. 

For our simulations, due to computational power limitations, we consider the case wherein the trapped ultracold atom gas has an order of magnitude equal to 3. Also, our initial states will be a superposition of four number states, which has the following explicit form:

\begin{eqnarray}
&&\left|\psi_{0}\right\rangle = \frac{1}{\sqrt{4}}\left(\left(\left|(N_{g,1})_1\right\rangle+\left|(N_{g,1})_2\right\rangle+\left|(N_{g,1})_3\right\rangle+\left|(N_{g,1})_4\right\rangle\right)+\left(\left|(N_{g,2})_1\right\rangle+\left|(N_{g,2})_2\right\rangle\right.\right.\nonumber\\
&&\left.\left.+\left|(N_{g,2})_3\right\rangle+\left|(N_{g,2})_4\right\rangle\right)+\left(\left|(N_{g,3})_1\right\rangle+\left|(N_{g,3})_2\right\rangle+\left|(N_{g,3})_3\right\rangle+\left|(N_{g,3})_4\right\rangle\right)\right)\nonumber\\
\label{initsysstate}
\end{eqnarray}

Here, the state $\left|(N_{g,j})_n\right\rangle$ is an eigenstate of the particle number operator $\hat{N}_{g,j}=\hat{a}^{\dagger}_{g,j}\hat{a}_{g,j}$ corresponding to the ground state of the system located at the $j$th node with particle number $(N_{g,j})_n$, so that $\hat{N}_{g,j}\left|(N_{g,j})_n\right\rangle=(N_{g,j})_{n}\left|(N_{g,2})_1\right\rangle$. The density matrix corresponding to the initial state will then have the form $\hat{\rho}_{0}=\left|\psi_{0}\right\rangle\left\langle\psi_{0}\right|$. 

\subsection{$\left\langle N_{1}(t)\right\rangle > \left\langle N_{2}(t)\right\rangle$ and $\left\langle N_{1}(t)\right\rangle > \left\langle N_{3}(t)\right\rangle$}

We first consider the case wherein most of the trapped ultracold atoms are initially loaded into the first trap, i. e. the trap where the first ground state is located, such that $\left\langle N_{1}(t)\right\rangle > \left\langle N_{2}(t)\right\rangle$ and $\left\langle N_{1}(t)\right\rangle > \left\langle N_{3}(t)\right\rangle$. For this case, the particle number eigenstates which are components of the initial state $\left|\psi_{0}\right\rangle$ have corresponding eigenvalues 
\begin{eqnarray}
&&(N_{g,1})_1 = 325, (N_{g,1})_2 = 330, (N_{g,1})_3 = 370, (N_{g,1})_4 = 375, (N_{g,2})_1 = 120, (N_{g,2})_2 = 130,\nonumber\\ 
&&(N_{g,2})_3 = 150, (N_{g,2})_4 = 200, (N_{g,3})_1 = 50, (N_{g,3})_2 = 75, (N_{g,3})_3 = 125, (N_{g,3})_4 = 150\nonumber\\
\label{numopeigenvalcase1} 
\end{eqnarray}
These eigenstates, in turn, will correspond to the particle number expectation values $\left\langle\hat{N}_{g,1}\right\rangle=350, \left\langle\hat{N}_{g,2}\right\rangle=150, \left\langle\hat{N}_{g,3}\right\rangle=100$ which all add up to a total particle number expectation value of $\left\langle\hat{N}\right\rangle=\left\langle\hat{N}_{g,1}\right\rangle+\left\langle\hat{N}_{g,2}\right\rangle+\left\langle\hat{N}_{g,3}\right\rangle=600$. In evolving this initial state, we set the coupling constant in the master equation used to evolve the state to have the value $A=9.00\times 10^-5$, and the characteristic timescale over which the state is evolved to have the value $\tau_E = \hbar/\Delta \epsilon = 0.1 s$, where $\Delta \epsilon = \epsilon_e -\epsilon_g$, with the state evolved over 100 timesteps. 

\begin{figure}[htb]
\includegraphics[width=0.5\columnwidth, height=0.2\textheight]{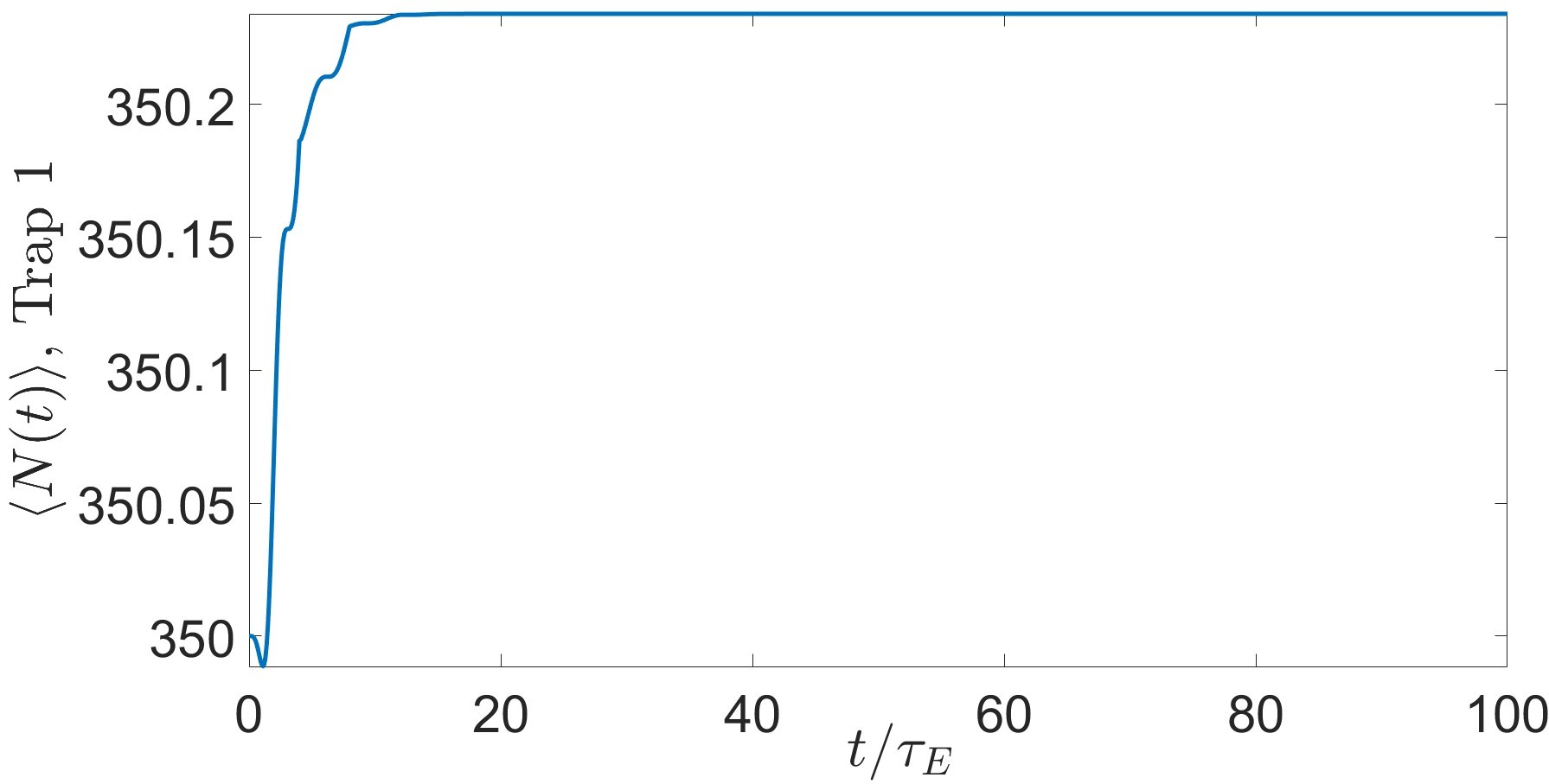}
\includegraphics[width=0.5\columnwidth, height=0.2\textheight]{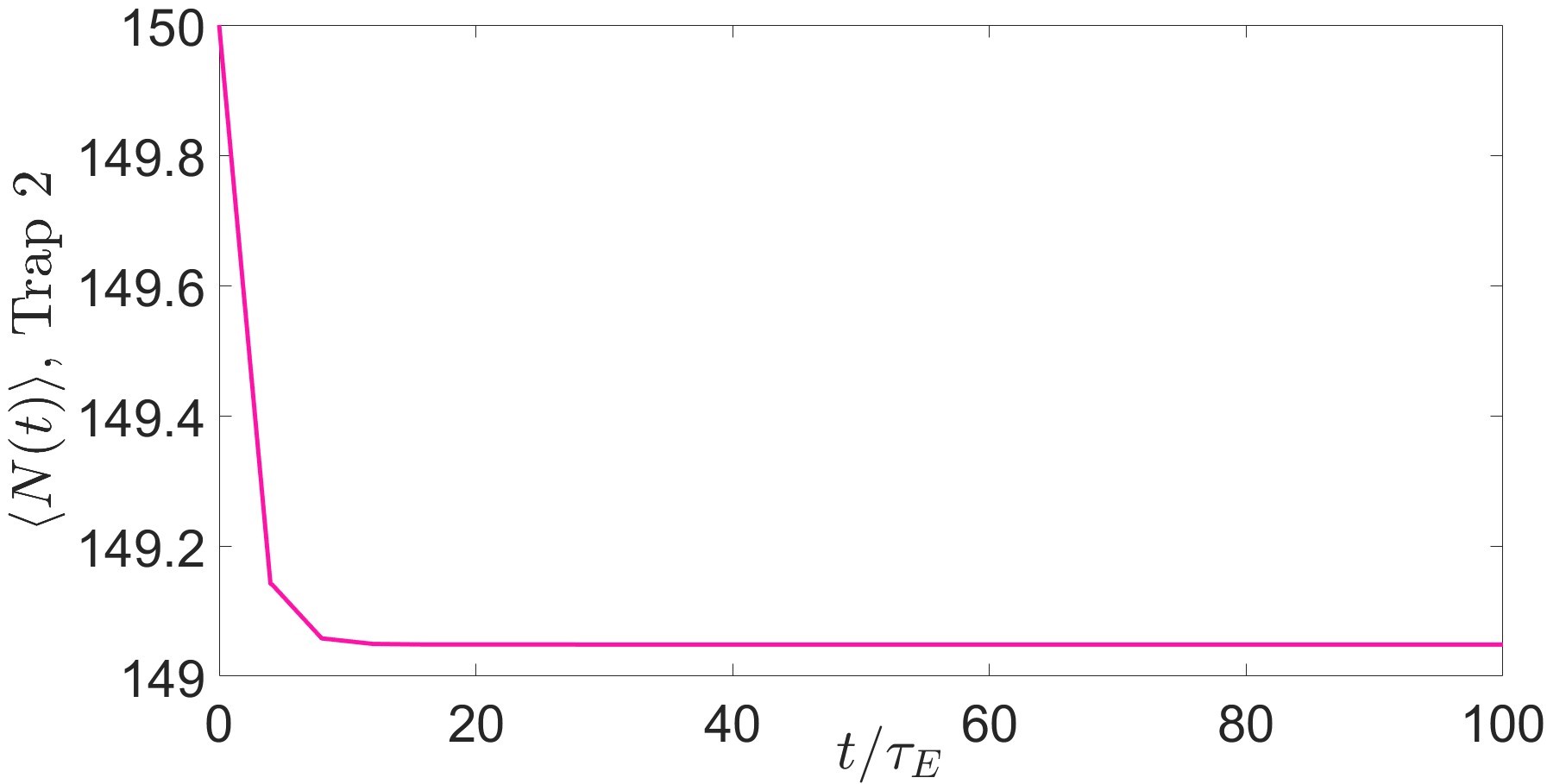}
\includegraphics[width=0.5\columnwidth, height=0.2\textheight]{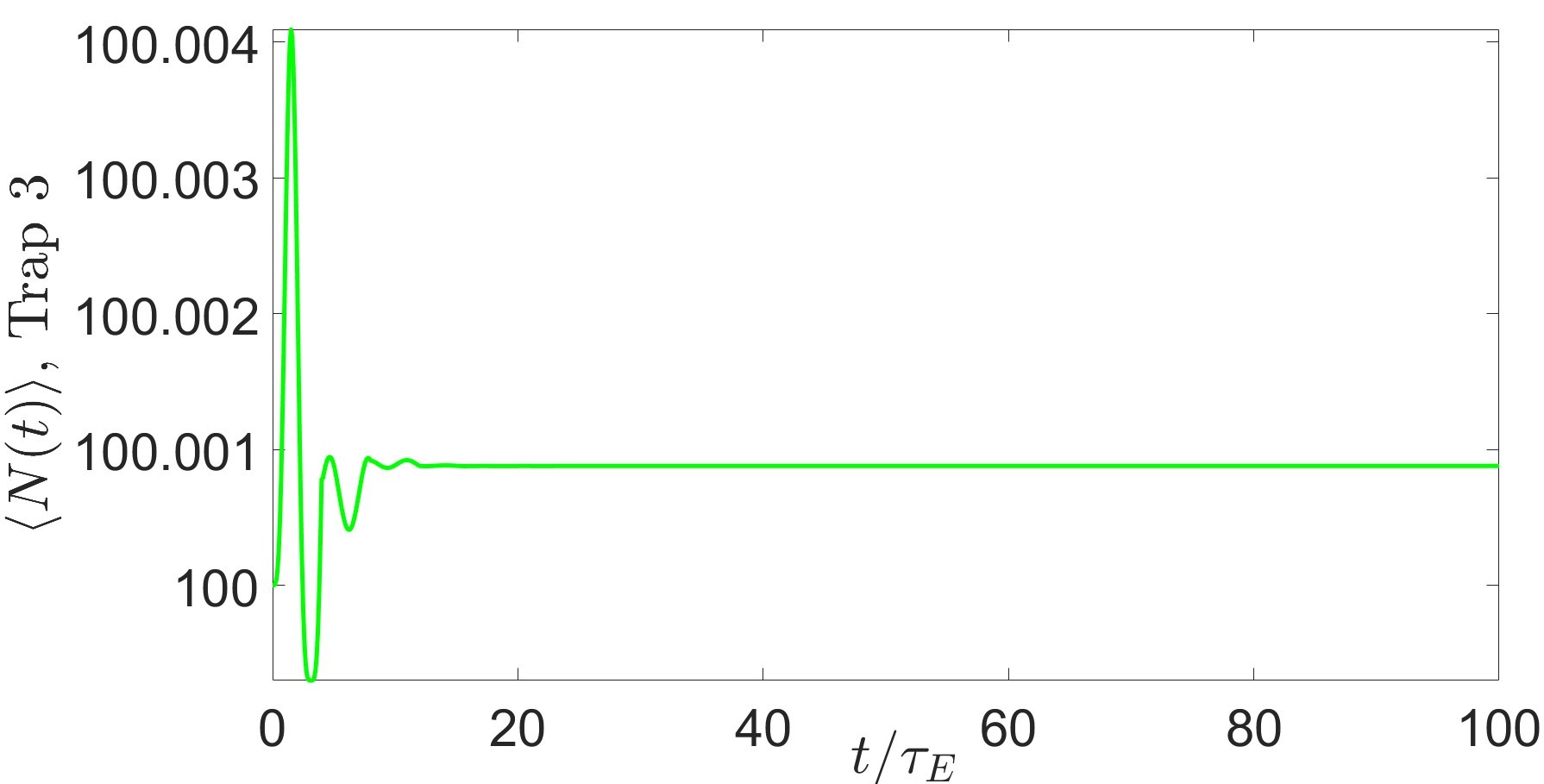}
\caption{\label{fig:numexpval1} Time evolution of the expectation values of the particle number $N(t)$ in the first (upper left), second (upper right) and third (lower left) harmonic traps containing the ground state of the system, calculated using a superposition of four particle number eigenstates per ground state location, with corresponding eigenvalues given in Eq. \ref{numopeigenvalcase1}.}
\end{figure}

As can be seen in Fig. \ref{fig:numexpval1}, for the given parameters of the system, each of the components of the ultracold atom gas trapped in the three harmonic potentials corresponding to the ground states of the system will evolve via a combination of their interaction with the background BEC and their stroboscopic coupling to the excited states of the system located in neighboring harmonic traps in such a way that the expectation value of the particle number of these ultracold atom gas components will approach a steady - state value over time. In particular, the expectation value of the number of ultracold atoms in the first harmonic trap will continuously increase until it reaches its steady - state value, while the expectation value of the number of ultracold atoms in the second harmonic trap will continuously decrease until it reaches its steady - state value. On the other hand, the expectation value of the number of ultracold atoms in the third harmonic trap oscillates until its steady - state value is attained. This dynamical behavior of the three components of the ultracold atom gas in their respective harmonic traps
can be seen for a wide range of values of the coupling constant $A$ and for a large set of combinations of particle number eigenstates $\left|(N_{g,j})_n\right\rangle$ that form the initial state. However, as can also be seen in Fig. \ref{fig:numexpval1}, the change in the initial number of particles in each trapped component of the ultracold atom gas is minimal, of the order of magnitude 1 for this example. But for a given set of particle number eigenstates $\left|(N_{g,j})_n\right\rangle$ and coupling constants $A$, this change in the particle number can be magnified to be of order of magnitude 2 or greater.

This is shown in this second case case wherein most of the trapped ultracold atoms are again initially loaded into the first trap, i. e. the trap where the first ground state is located. But for this case, the particle number eigenstates which are components of the initial state $\left|\psi_{0}\right\rangle$ have corresponding eigenvalues 
\begin{eqnarray}
&&(N_{g,1})_1 = 425, (N_{g,1})_2 = 426, (N_{g,1})_3 = 575, (N_{g,1})_4 = 574, (N_{g,2})_1 = 20, (N_{g,2})_2 = 25,\nonumber\\ 
&&(N_{g,2})_3 = 80, (N_{g,2})_4 = 70, (N_{g,3})_1 = 20, (N_{g,3})_2 = 25, (N_{g,3})_3 = 80, (N_{g,3})_4 = 75\nonumber\\
\label{numopeigenvalcase2} 
\end{eqnarray}
These eigenstates, in turn, will correspond to the particle number expectation values $\left\langle\hat{N}_{g,1}\right\rangle=500, \left\langle\hat{N}_{g,2}\right\rangle=50, \left\langle\hat{N}_{g,3}\right\rangle=50$ which all add up to a total particle number expectation value of $\left\langle\hat{N}\right\rangle=\left\langle\hat{N}_{g,1}\right\rangle+\left\langle\hat{N}_{g,2}\right\rangle+\left\langle\hat{N}_{g,3}\right\rangle=600$. In evolving this initial state, we set the coupling constant in the master equation used to evolve the state to have the value $A=1.00\times 10^{-4}$, and the characteristic timescale over which the state is evolved to have the value $\tau_E = 0.1$, with the state evolved over 100 timesteps.

\begin{figure}[htb]
\includegraphics[width=0.5\columnwidth, height=0.2\textheight]{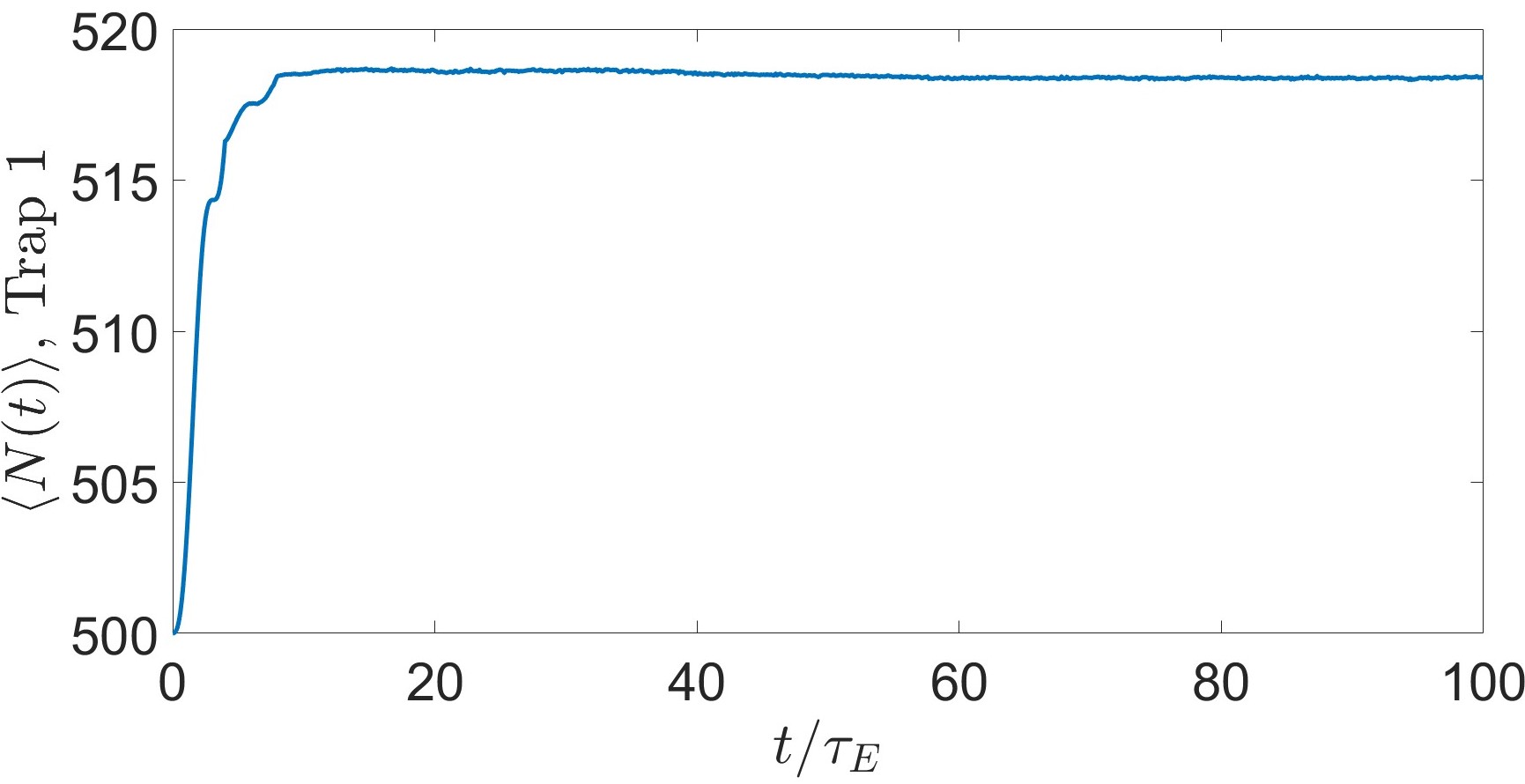}
\includegraphics[width=0.5\columnwidth, height=0.2\textheight]{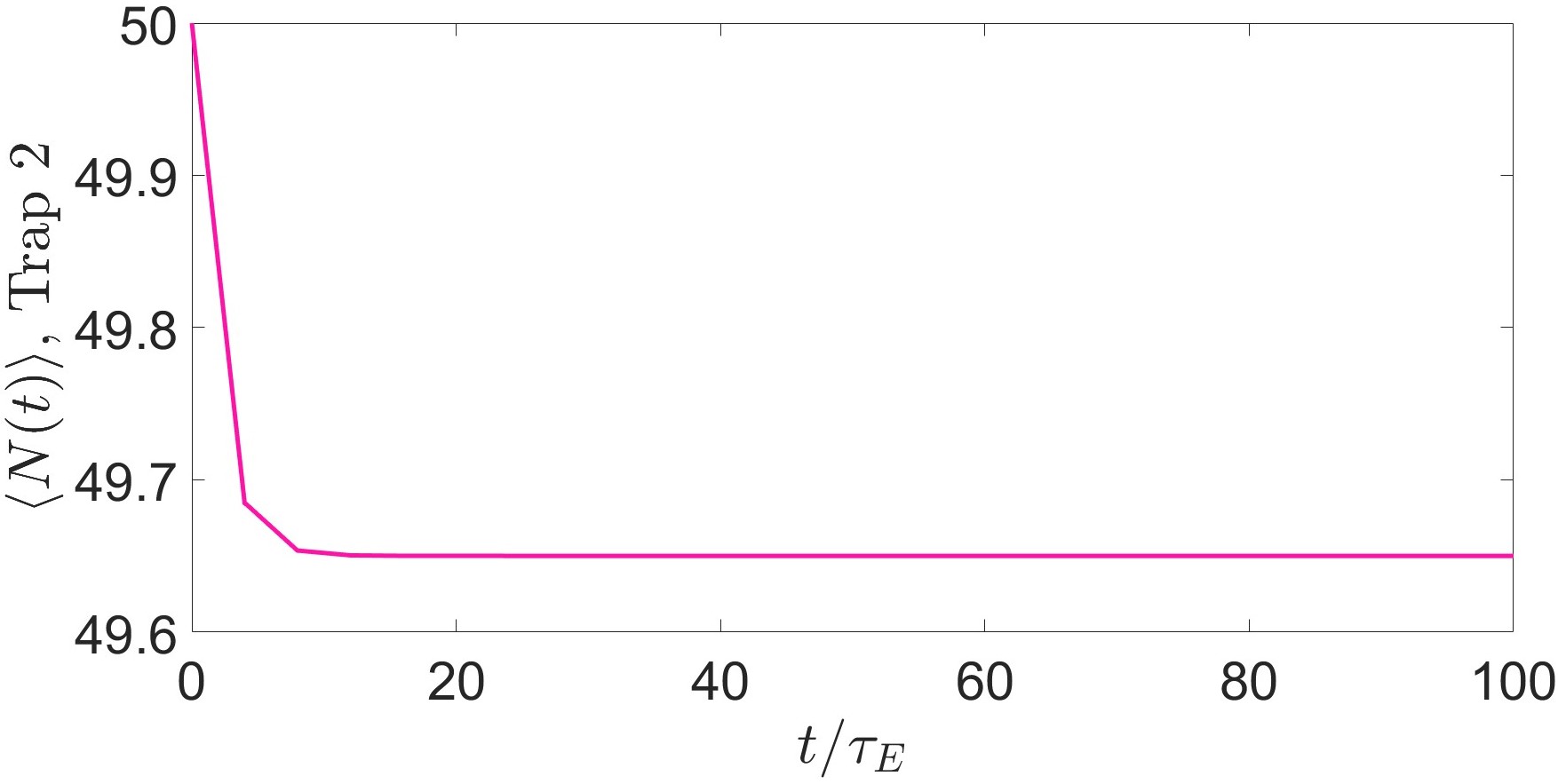}
\includegraphics[width=0.5\columnwidth, height=0.2\textheight]{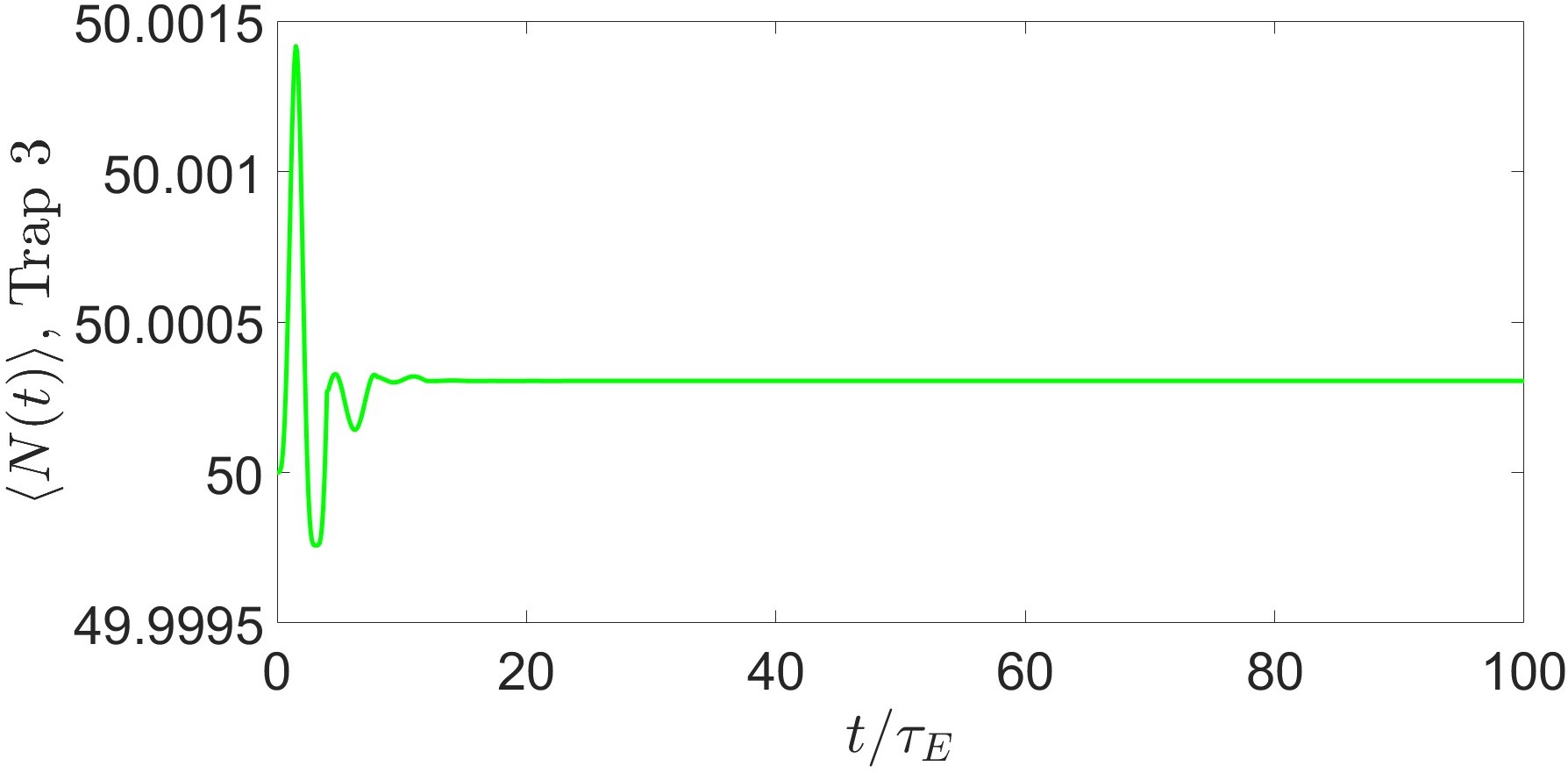}
\caption{\label{fig:numexpval2} Time evolution of the expectation values of the particle number $N(t)$ in the first (upper left), second (upper right) and third (lower left) harmonic traps containing the ground state of the system, calculated using a superposition of four particle number eigenstates per ground state location, with corresponding eigenvalues given in Eq. \ref{numopeigenvalcase2}.}
\end{figure}

As can be seen in Fig. \ref{fig:numexpval2}, for this combination of parameters, we can see that the number of particles in the first trap corresponding to the ground state of the system has increased by an order of magnitude of 2. However, we see a negligible change in the number of particles in the other two traps corresponding to the system's ground state. This then suggests that, due to the interaction with the background BEC, atoms from the BEC have been scattered into the trapped ultracold atom gas, mixing with them and becoming part of the system. One possible mechanism that can facilitate such scattering from the background BEC into the trapped ultracold atom gas is hyperfine scattering between background BEC atoms and the trapped ultracold gas's atoms. This may not be a problem if both the background BEC and the trapped ultracold atom gas are composed of atoms of the same species, but in cases wherein the background BEC and the trapped ultracold atom gas are composed of atoms of different species, such intermixing of atoms may present a problem especially if the trapped ultracold atom gas needs to be homogeneous, as is the case if the trapped ultracold atom gas will eventually undergo Bose - Einstein condensation. Hence, great care must be exercised in selecting the particle number eigenstates that will be used to form the initial state describing the trapped ultracold atom gas. 

\begin{figure}[htb]
\includegraphics[width=1.1\columnwidth, height=0.4\textheight]{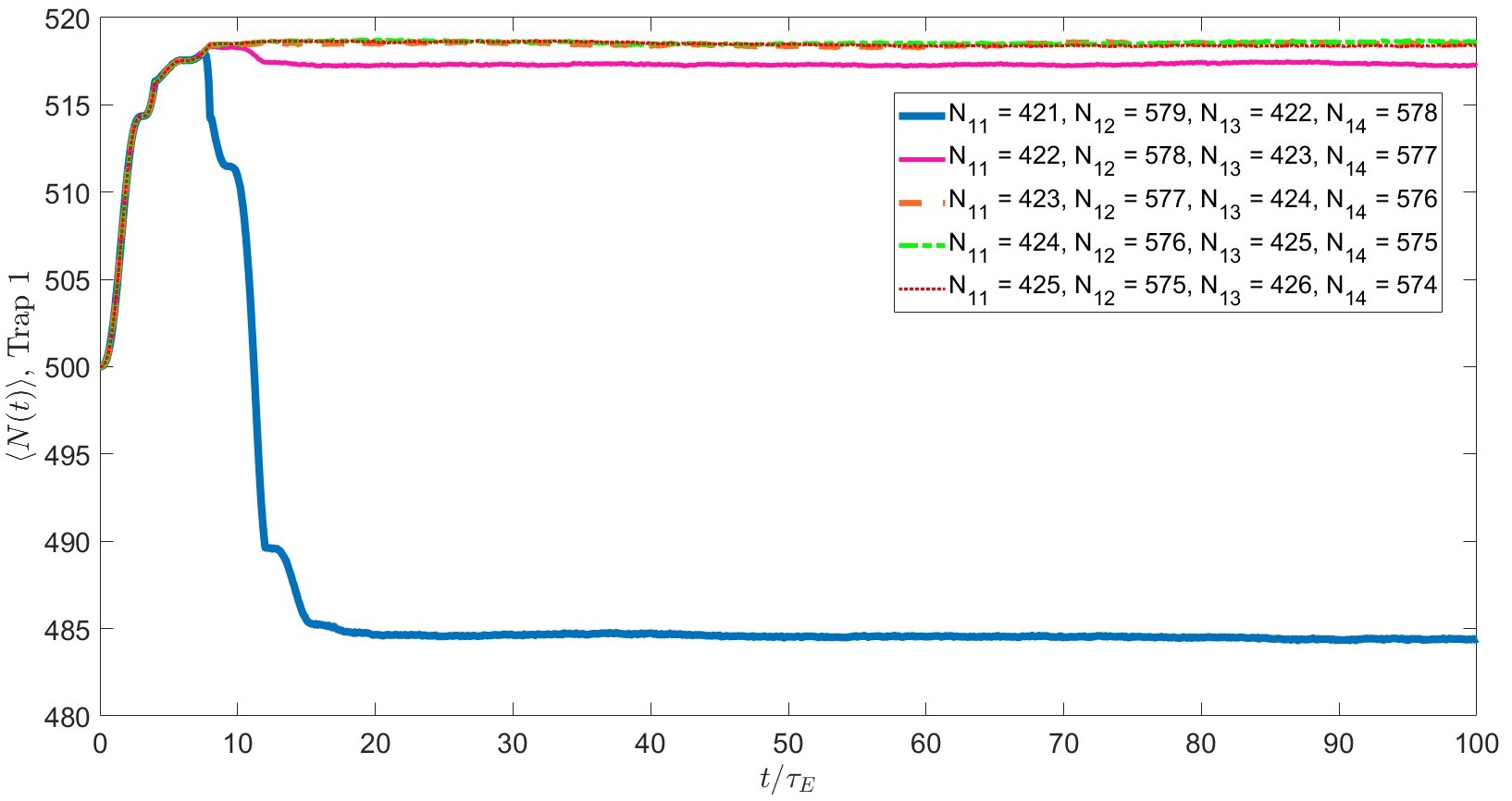}
\caption{\label{fig:numexpval3} Time evolution of the expectation values of the particle number $N(t)$ in the first  harmonic trap containing the ground state of the system, calculated using different superpositions of four particle number eigenstates per ground state location, with corresponding eigenvalues given in the figure. For this figure, $A = 1.00\times 10^-4$.}
\end{figure}

To further underscore the importance of our choice of particle number eigenstates that will form our initial state, let us consider the following figure, which shows how the expectation value of the particle number in trap 1 varies for different combinations of particle number eigenstates used to construct the component of the initial state corresponding to trap 1. As we can see in this figure, it is possible to construct an initial state wherein the particle number expectation value in trap 1 first increases, then decreases over time until it attains its steady - state value. However, the order of magnitude of the particle number decrease in this case is around 2, which is a relatively large amount. On the other hand, a small variation in the component particle number eigenstates for this initial state will result in large changes in the dynamical behavior of the system. As shown in the figure, if we change the initial state in trap 1 from 
\begin{equation}
\frac{1}{\sqrt{4}}\left(\left|(N_{g,1})_1 = 421\right\rangle+\left|(N_{g,1})_2 = 579\right\rangle+\left|(N_{g,1})_3 = 422\right\rangle+\left|(N_{g,1})_4 = 578\right\rangle\right)
\end{equation} 
to
\begin{equation}
\frac{1}{\sqrt{4}}\left(\left|(N_{g,1})_1 = 422\right\rangle+\left|(N_{g,1})_2 = 578\right\rangle+\left|(N_{g,1})_3 = 423\right\rangle+\left|(N_{g,1})_4 = 577\right\rangle\right)
\end{equation}
then the initial state suddenly changes from a state whose particle number expectation value decreases by an order of magnitude of 2 as it evolves towards its steady state value to a state whose particle number expectation value increases by the same order of magnitude as it evolves over time towards its steady state value. At the same time, any further changes to the initial state will not result in a noticeable change in the manner by which it evolves over time, as can be seen in figure \ref{fig:numexpval3}, wherein the eigenvalues corresponding to the component particle number eigenstates of the initial state in trap 1 are changed by a value equal to 1. As the eigenvalues corresponding to the first and third component particle number eigenstates increase (with a corresponding decrease in the eigenvalues corresponding to the second and fourth component particle number eigenstates), the initial state in trap 1 will eventually evolve in such a way that the expectation value of its particle number increases over time and attains a definite steady - state value. However, the magnitude of this steady - state value remains constant even as the eigenvalues corresponding to the first and third component particle number eigenstates of the initial state in trap 1 continue to increase in magnitude. This case then shows that it is possible to find a set of parameters for the initial state of the system and evolve it in such a way that not only will the expectation value of the number of particles in trap 1 attain a steady - state value irrespective of the combination of particle number eigenstates that form the component of the system's initial state in trap 1, but also that this steady - state expectation value is greater than its initial value. 

\begin{figure}[htb]
\includegraphics[width=1.1\columnwidth, height=0.4\textheight]{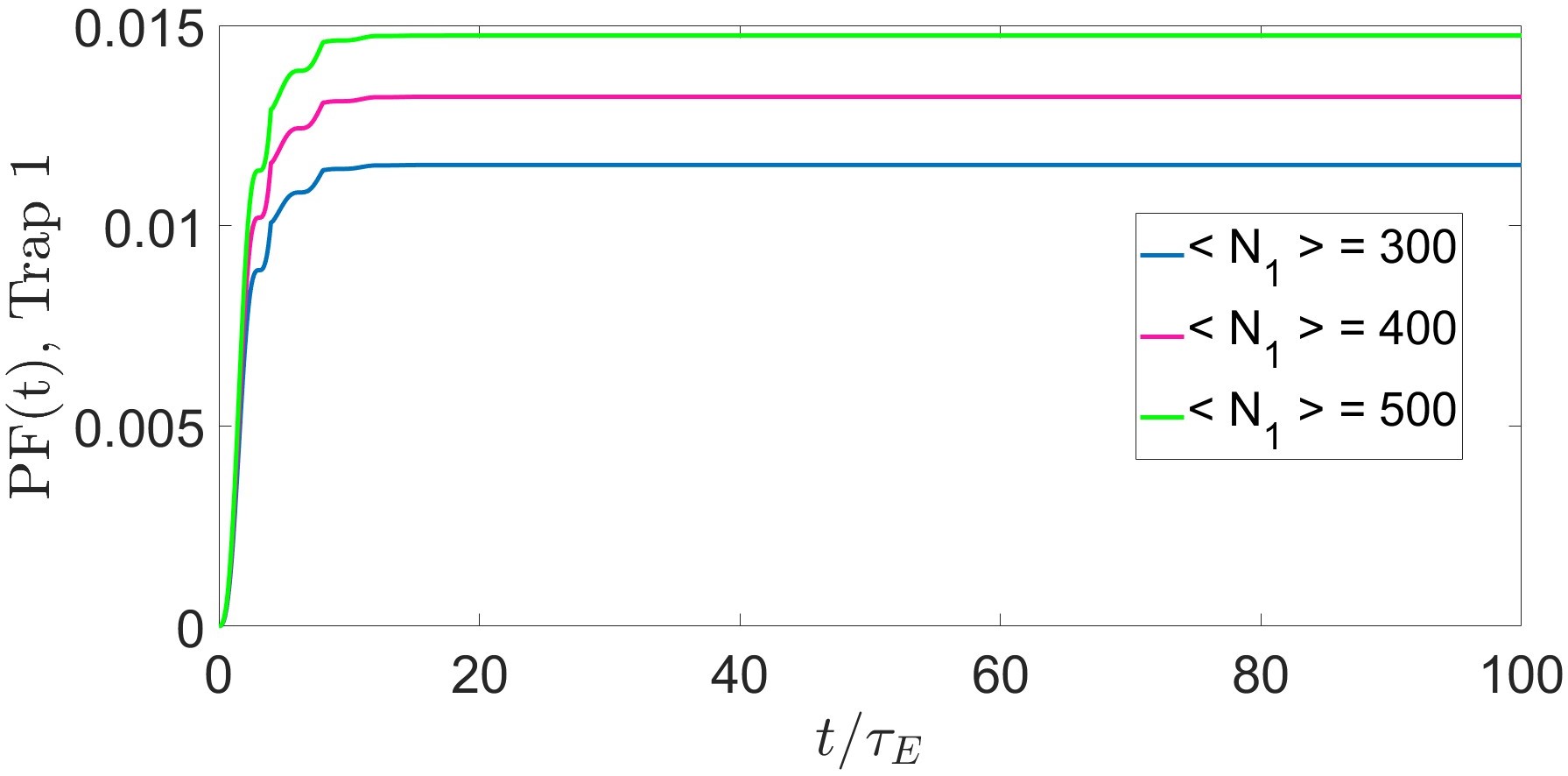}
\caption{\label{fig:condfrac1} Time evolution of the particle number fraction $PF(t)$ of the ultracold atom system's component in trap 1, with varying initial values of the particle number expectation value as indicated in the figure.}
\end{figure}

Now let us consider what happens to the time evolution of the initial state as the expectation value of the number of particles of the initial state in trap 1, $\left\langle N_1\right\rangle$ increases, when most of the particles in the trapped ultracold atom gas are initially in trap 1. To do this, we need to compute the particle number fraction, which we define as follows:

\begin{equation}
PF(t)=\frac{\left\langle N_{j}(t)\right\rangle - \left\langle N_{j}\right\rangle}{\left\langle N_{j}\right\rangle}
\label{particlefraction}
\end{equation}
Here, $\left\langle N_{j}(t)\right\rangle$ is the expectation value of the particle number in trap j, $j=1,2,3$, of the time - evolved state at the instant of time $t$, while $\left\langle N_{j}\right\rangle$ is the expectation value of the particle number in trap j for the initial state. As shown in Fig. \ref{fig:condfrac1}, as $\left\langle N_1\right\rangle$ increases, the particle number fraction in trap 1 also increases for any given instant of time $t$, and will eventually approach a steady - state value. This signifies that the greater the expectation value of the initial state of the system in trap 1, the greater the number of atoms from the background BEC that can mix with those from the ultracold atom gas in this trap, consequently increasing the expectation value of the trapped ultracold atom gas's particle number over time. 

\begin{figure}[htb]
\includegraphics[width=1.1\columnwidth, height=0.4\textheight]{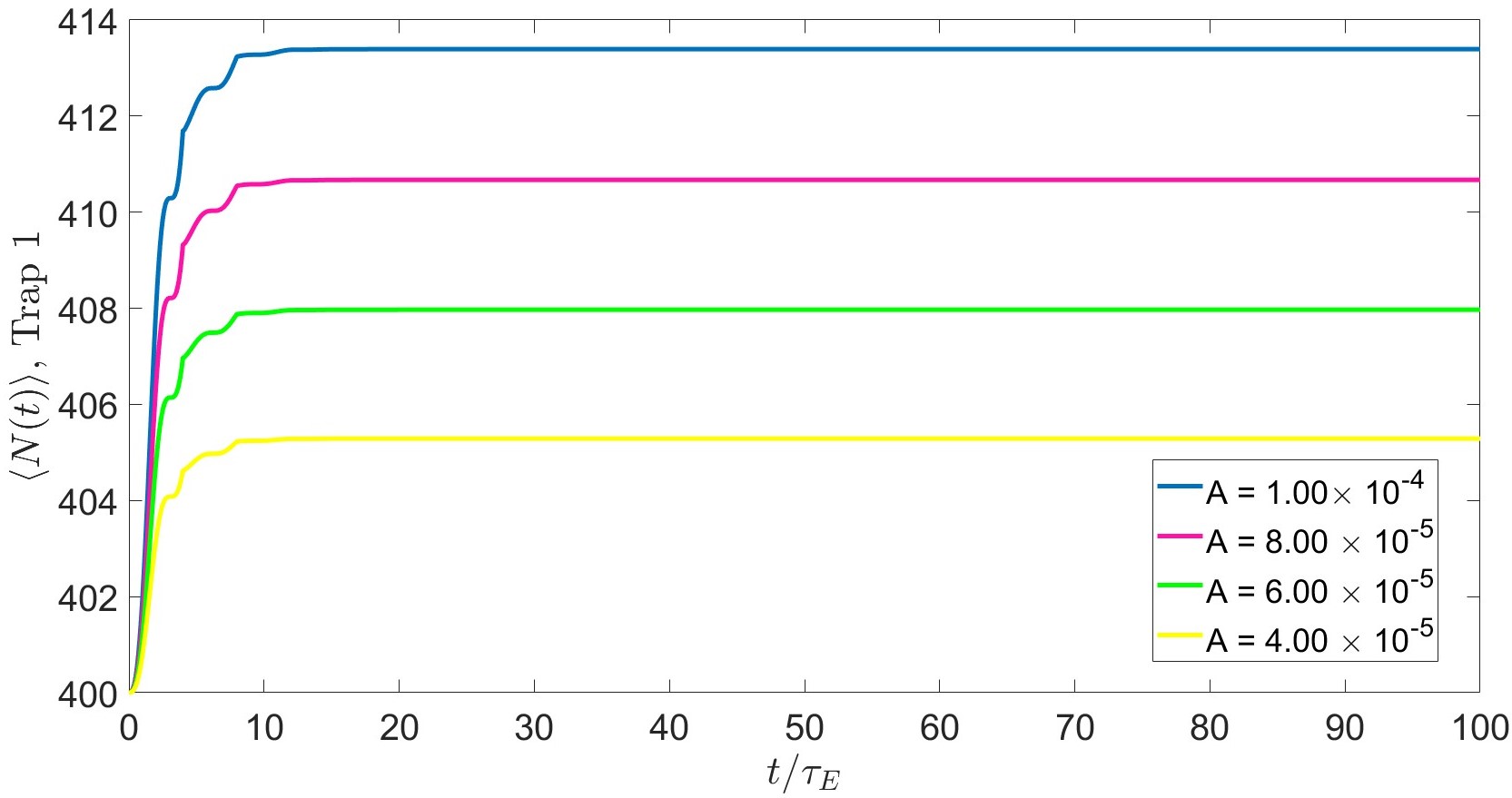}
\caption{\label{fig:numexpvalvaryA} Time evolution of the expectation values of the particle number $N(t)$ in the first  harmonic trap containing the ground state of the system, calculated with varying values of the coupling constant $A$ in the master equation, with these values of $A$ indicated in the figure.}
\end{figure}

Finally, let us see what happens when we adjust the coupling strength between the trapped ultracold atom system and the background BEC, which is done by varying the value of the coupling constant $A$ in the master equation. As shown in Fig. \ref{fig:numexpvalvaryA}, varying the magnitude of the coupling constant will have no effect on the manner by which the initial state in Trap 1 of the system will evolve over time; it will still evolve in such a way that $\left\langle N(t)\right\rangle$ will attain a steady - state value. However, as $A$ decreases, so too will the steady - state value of $\left\langle N(t)\right\rangle$. This signifies that weakening the coupling between the trapped ultracold atom system and the background BEC will also reduce the number of atoms from the background BEC mixing with the trapped ultracold atom gas if most of the atoms in the trapped ultracold gas are located in trap 1.

\subsection{$\left\langle N_{2}(t)\right\rangle > \left\langle N_{1}(t)\right\rangle$ and $\left\langle N_{2}(t)\right\rangle > \left\langle N_{3}(t)\right\rangle$}

Let us now consider the case where most of the particles in the ultracold atom gas are loaded in trap 2, such that $\left\langle N_{2}(t)\right\rangle > \left\langle N_{1}(t)\right\rangle$ and $\left\langle N_{2}(t)\right\rangle > \left\langle N_{3}(t)\right\rangle$. It is important to note that, from Eq. \ref{jumpop}, creation and annihilation operators corresponding to the location of trap 2 appear twice in the jump operator corresponding to the driven - dissipative dynamics of the system. This can be explained by the fact that, as shown in figure \ref{fig:harmonictraparray}, there are two excited energy levels, both of the same magnitude, that are coupled to the ground state energy level located in trap 2. As such, we expect that if the particle number expectation value at trap 2 of the initial state of the system is greater than the corresponding particle number expectation values at the other two traps for the same initial state, the dynamics of the system will be different from the case wherein the particle number expectation value at trap 1 of the system's initial state is greater than for the other two. This is, indeed the case when the particle number expectation values for the component particle number eigenstates of the initial state of the system have the following values:
\begin{eqnarray}
&&(N_{g,1})_1 = 120, (N_{g,1})_2 = 130, (N_{g,1})_3 = 150, (N_{g,1})_4 = 200, (N_{g,2})_1 = 325, (N_{g,2})_2 = 330,\nonumber\\ 
&&(N_{g,2})_3 = 370, (N_{g,2})_4 = 375, (N_{g,3})_1 = 50, (N_{g,3})_2 = 75, (N_{g,3})_3 = 150, (N_{g,3})_4 = 125\nonumber\\
\label{numopeigenvalcase3} 
\end{eqnarray}
For this case, the corresponding particle number expectation values in each trap are $\left\langle\hat{N}_{g,1}\right\rangle=150, \left\langle\hat{N}_{g,2}\right\rangle=350, \left\langle\hat{N}_{g,3}\right\rangle=100$ which all add up to a total particle number expectation value of $\left\langle\hat{N}\right\rangle=\left\langle\hat{N}_{g,1}\right\rangle+\left\langle\hat{N}_{g,2}\right\rangle+\left\langle\hat{N}_{g,3}\right\rangle=600$.

\begin{figure}[htb]
\includegraphics[width=0.5\columnwidth, height=0.2\textheight]{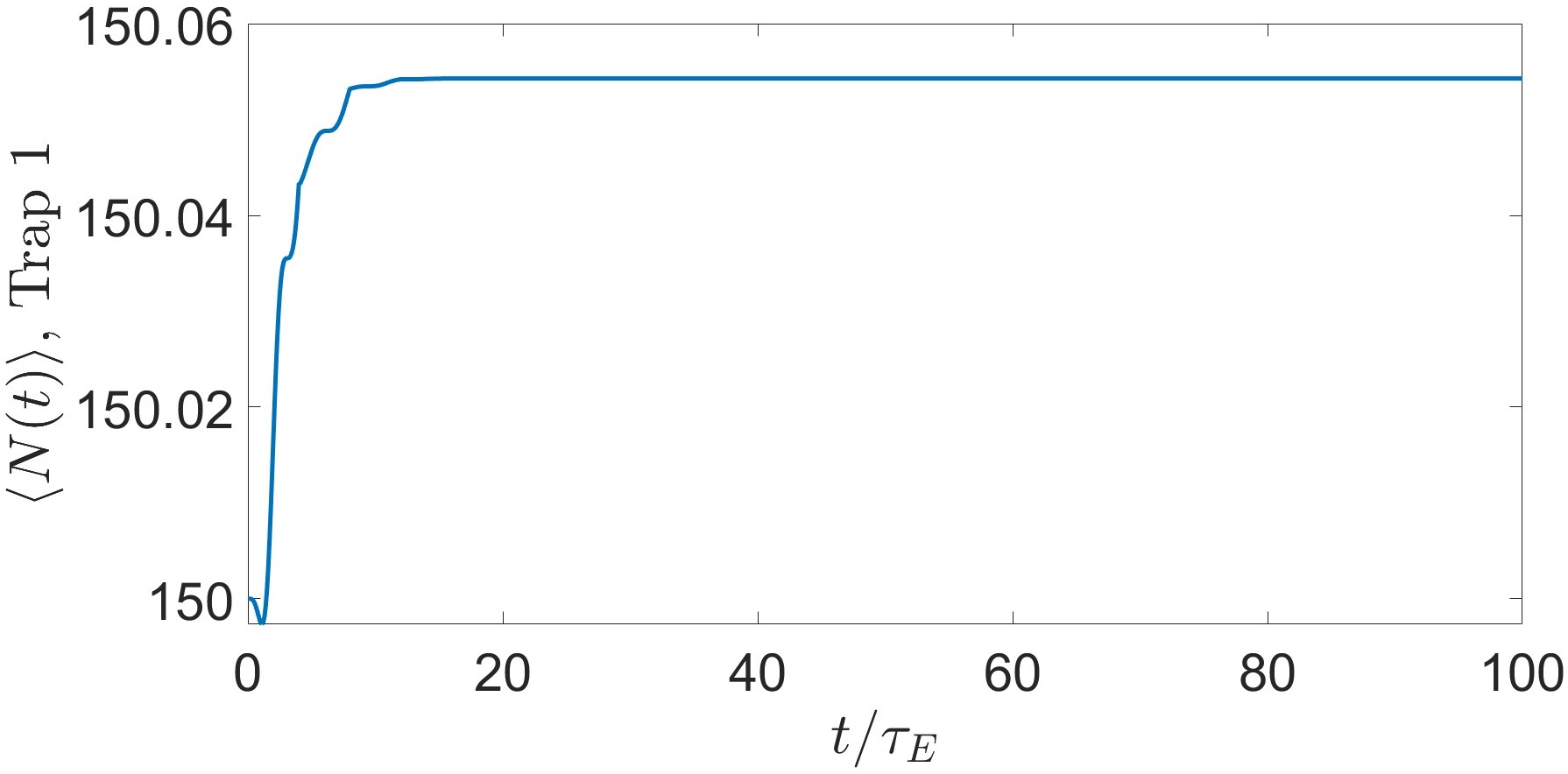}
\includegraphics[width=0.5\columnwidth, height=0.2\textheight]{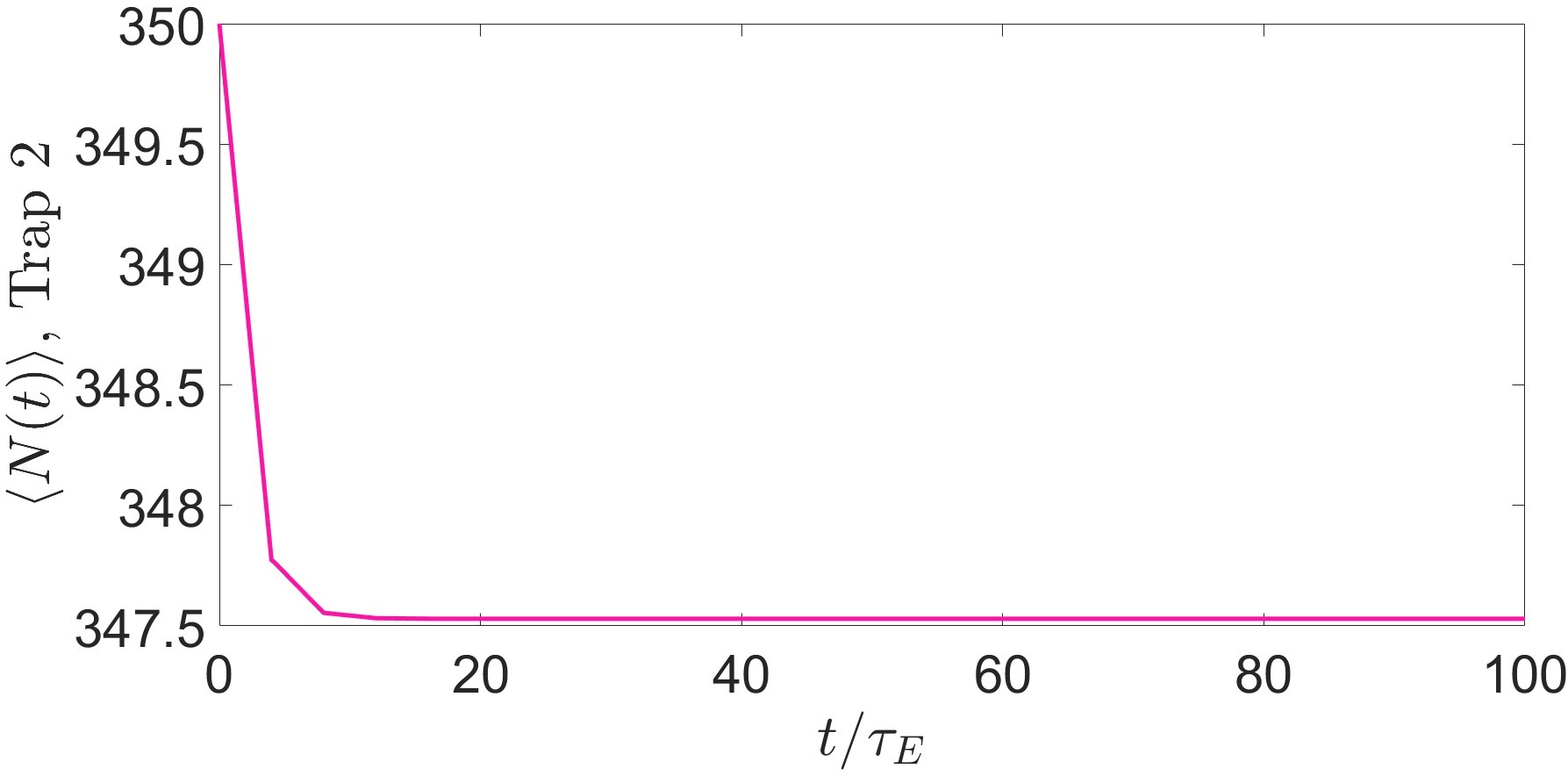}
\includegraphics[width=0.5\columnwidth, height=0.2\textheight]{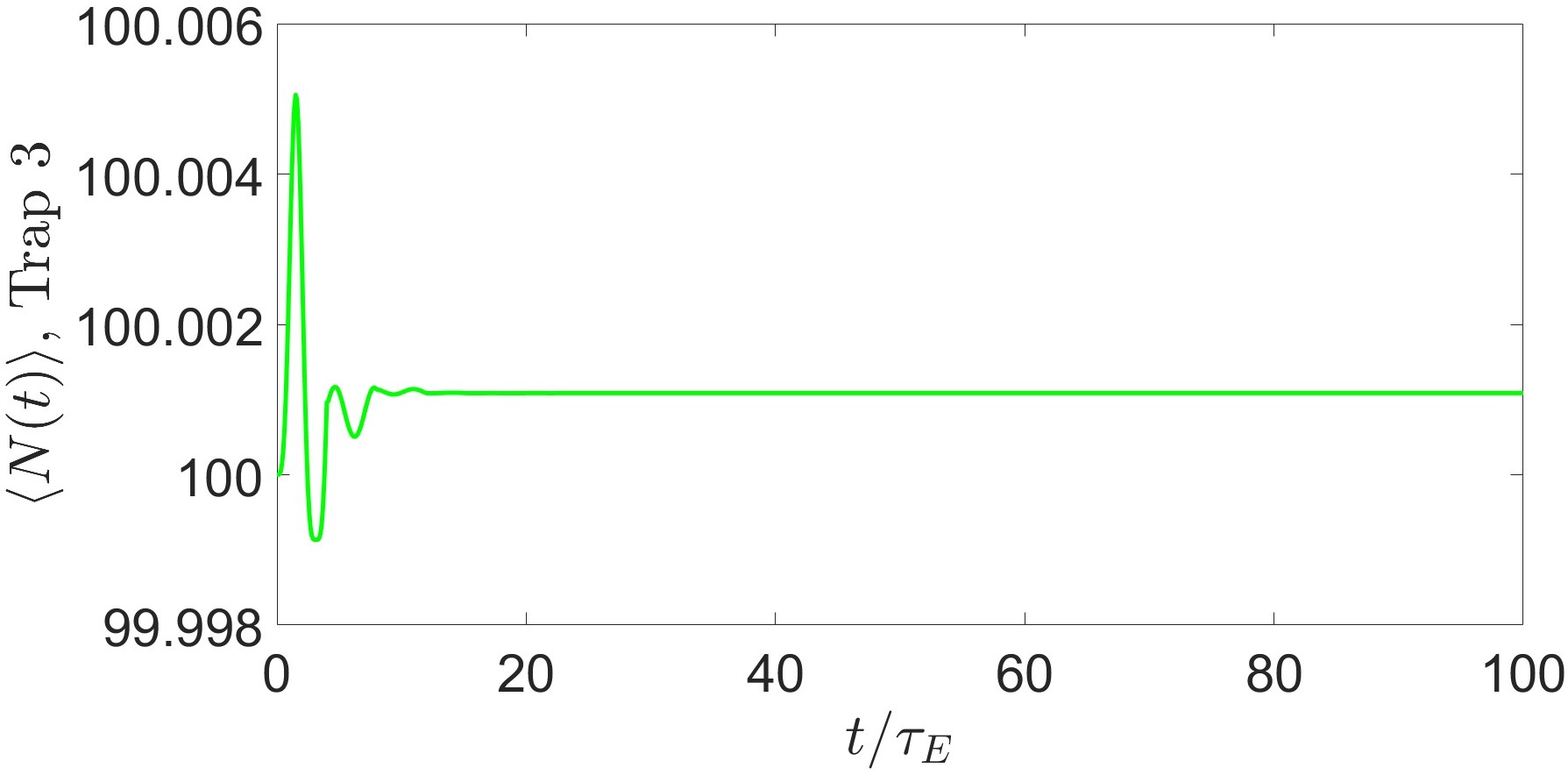}
\caption{\label{fig:numexpval4} Time evolution of the expectation values of the particle number $N(t)$ in the first (upper left), second (upper right) and third (lower left) harmonic traps containing the ground state of the system, calculated using a superposition of four particle number eigenstates per ground state location, with corresponding eigenvalues given in Eq. \ref{numopeigenvalcase3}.}
\end{figure}

Now the first thing that we should note for this case is that the coupling constant $A$ of the master equation used to evolve this state is larger by an order of magnitude equal to 1 compared to the coupling constant $A$ of the master equation used to evolve the first two initial states. In particular, for this case, the coupling constant $A$ has a value $A=1.00\times 10^{-3}$. This in itself already signifies that the dynamics for this case is already different from the previous two, due to the coupling being stronger for this case. It should be noted that if this magnitude of the coupling constant were used for the previous two cases, with all other parameters for those cases remaining unchanged, the dynamics of the time - evolved initial state will be very different from what we have seen in Fig. \ref{fig:numexpval4}. At the same time, if one is to compare the dynamics of the time - evolved initial state from Figs. \ref{fig:numexpval1} and \ref{fig:numexpval4}, we find that qualitatively, they may look similar, with the particle number expectation value in trap 1 steadily increasing until the steady - state value is reached; the particle number expectation value in trap 2 steadily decreasing until the steady - state value is reached; and the particle number expectation value in trap 3 oscillating until its steady - state value is reached. However, there is a fundamental difference in the magnitude of increase and decrease in particle number expectation values for traps 1 and 2, respectively, for these two different initial states. In particular, for trap 1, the magnitude of increase of the particle number expectation value is much greater for the case where $\left\langle N_{1}(t)\right\rangle > \left\langle N_{2}(t)\right\rangle$ than in the case where $\left\langle N_{1}(t)\right\rangle < \left\langle N_{2}(t)\right\rangle$, while for trap 2, the magnitude of decrease of the particle number expectation value is much greater for the case where $\left\langle N_{1}(t)\right\rangle < \left\langle N_{2}(t)\right\rangle$ than in the case where $\left\langle N_{1}(t)\right\rangle > \left\langle N_{2}(t)\right\rangle$. Furthermore, the order of magnitude of the difference between the particle number expectation values in trap 1 for the cases $\left\langle N_{1}(t)\right\rangle > \left\langle N_{2}(t)\right\rangle$ and $\left\langle N_{1}(t)\right\rangle < \left\langle N_{2}(t)\right\rangle$ is much less than the order of magnitude of the difference between the particle number expectation values in trap 2 for the cases $\left\langle N_{1}(t)\right\rangle > \left\langle N_{2}(t)\right\rangle$ and $\left\langle N_{1}(t)\right\rangle < \left\langle N_{2}(t)\right\rangle$.

\begin{figure}[htb]
\includegraphics[width=1.1\columnwidth, height=0.4\textheight]{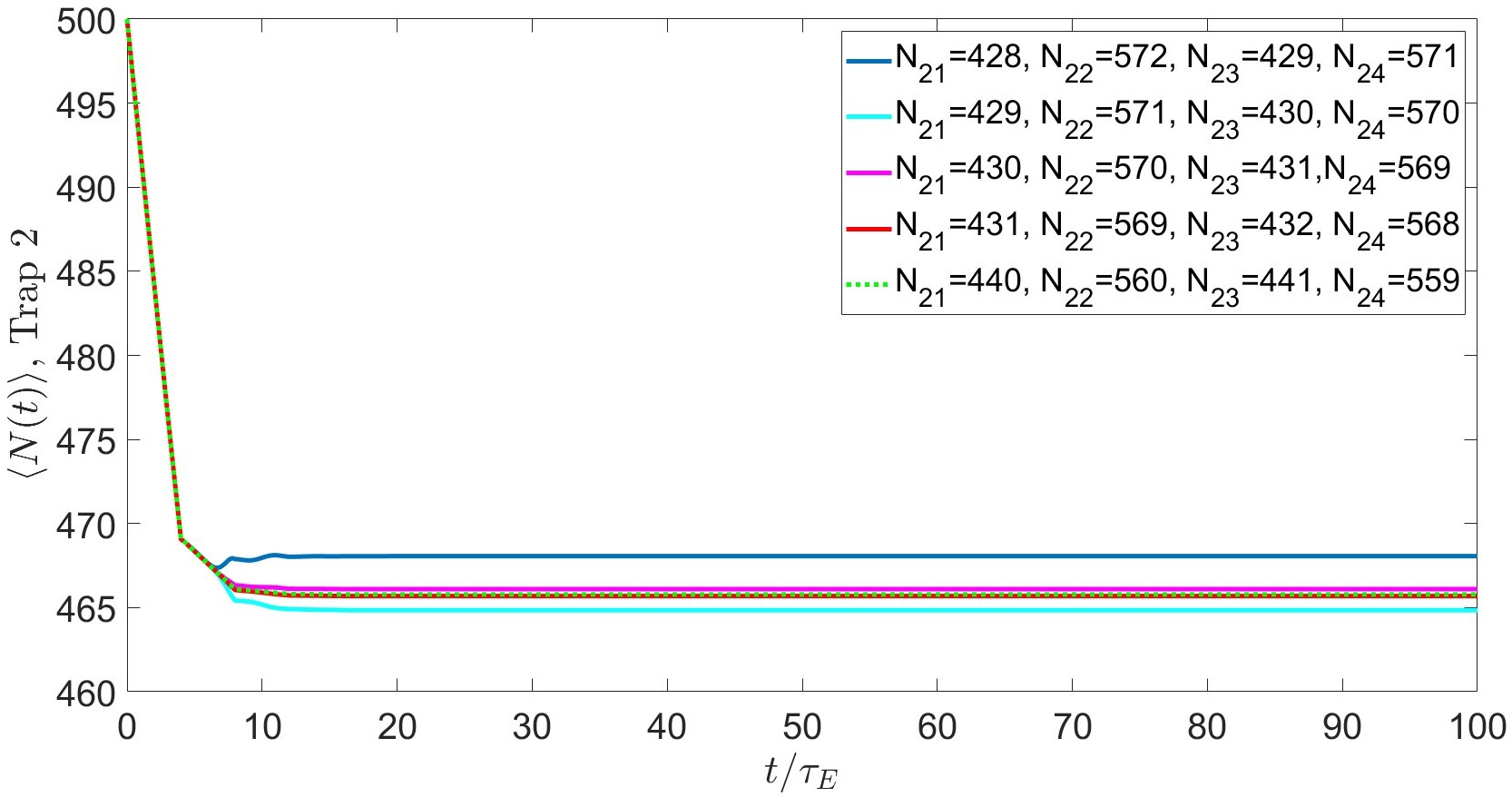}
\caption{\label{fig:numexpval5} Time evolution of the expectation values of the particle number $N(t)$ in the second harmonic trap containing the ground state of the system, calculated using different superpositions of four particle number eigenstates per ground state location, with corresponding eigenvalues given in the figure.}
\end{figure}

Now let us consider what happens when we vary the particle number eigenstates that form the initial state of the system for this case where most of the particles are to be found in Trap 2. For the case under consideration, the particle number expectation values in each trap are given as $\left\langle N_{1}\right\rangle = 100, \left\langle N_{2}\right\rangle = 500, \left\langle N_{3}\right\rangle = 100$. As shown in Fig. \ref{fig:numexpval5}, the expectation value of the number of particles in Trap 2 for this case will vary as we change the component particle number eigenstates, and their corresponding eigenvalues, that form the initial state of the system in Trap 2. However, despite this variation in the component particle number eigenstates, there is no observed qualitative change in how the trapped ultracold atom system's component in Trap 2 evolves over time, as can be seen in Fig.\ref{fig:numexpval5}. In particular, it can be seen that the magnitude of the particle number expectation value in Trap 2 oscillates, eventually settling towards a central value when $\left(N_{g,2}\right)_1 = 440, \left(N_{g,2}\right)_2 = 560, \left(N_{g,2}\right)_3 = 441, \left(N_{g,2}\right)_4 = 559$. However, the manner in which the time - evolved initial state in Trap 2 evolves towards this steady - state remains the same, whereby the expectation value of its particle number continuously decreases until it reaches the steady - state value. At the same time, for this case, it can be seen that even if the eigenvalues corresponding to the first and third component particle number eigenstates for the initial state in Trap 2 increase in magnitude, the steady - state attained by the time - evolved state in this location will remain the same.

\begin{figure}[htb]
\includegraphics[width=1.1\columnwidth, height=0.4\textheight]{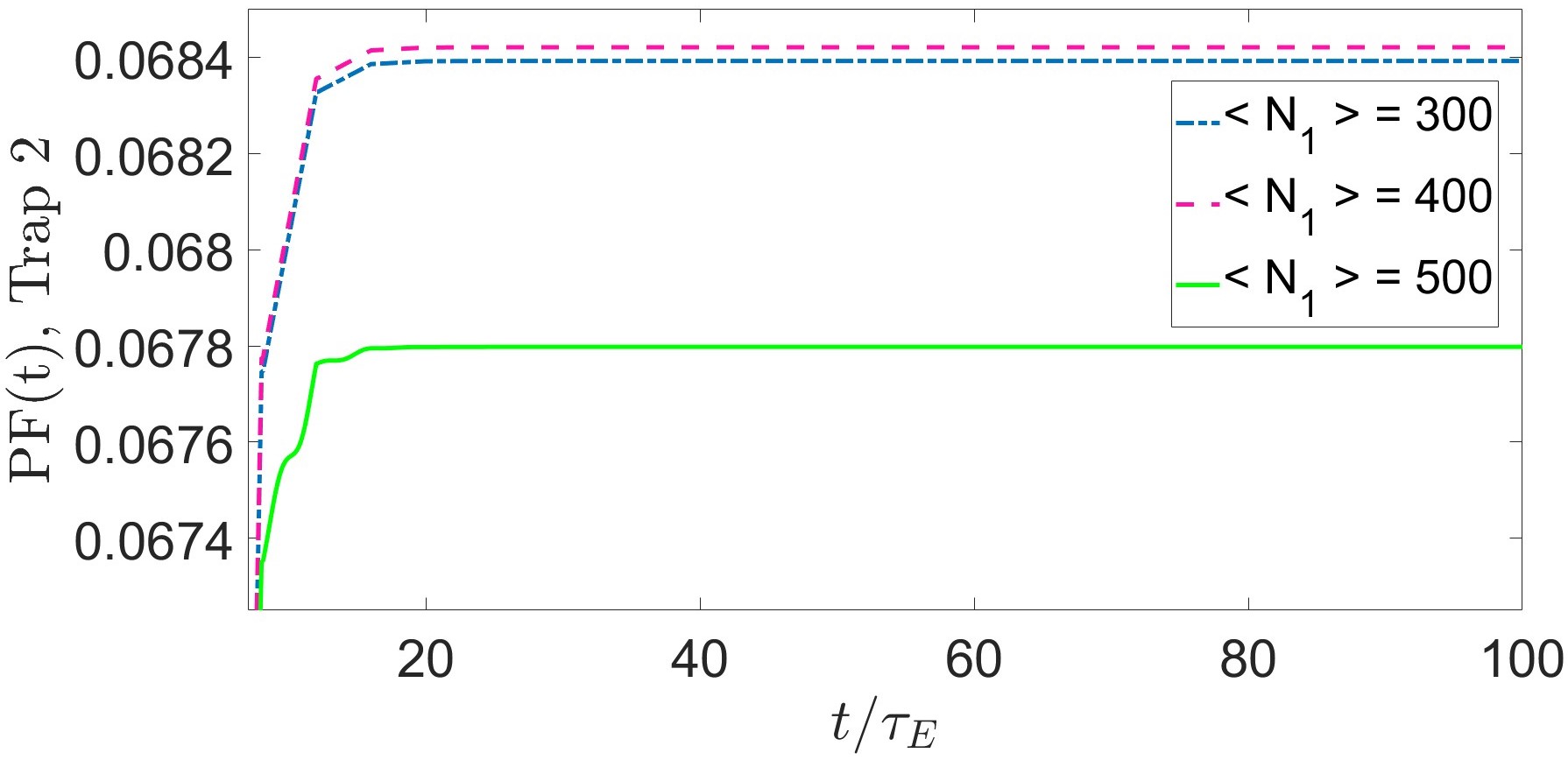}
\caption{\label{fig:condfrac2} Time evolution of the particle number fraction $PF(t)$ of the ultracold atom system's component in trap 2, with varying initial values of the particle number expectation value as indicated in the figure.}
\end{figure}

Looking at how the particle number fraction $PF(t)$ in trap 2 as defined in Eq. \ref{particlefraction} will vary for this case as we increase the expectation number of the particle number of the initial state in trap 2, we find that, as shown in Fig. \ref{fig:condfrac2}, $PF(t)$ behaves differently from the previous case. In particular, for this case, the steady - state value of $PF(t)$ first increases slightly as $\left\langle N_{2}\right\rangle$ increases from $\left\langle N_{2}\right\rangle = 300$ to $\left\langle N_{2}\right\rangle = 400$, then decreases significantly as $\left\langle N_{2}\right\rangle$ increases from $\left\langle N_{2}\right\rangle = 400$ to $\left\langle N_{2}\right\rangle = 500$. However, for all three cases, just as in the previous case, $PF(t)$ will again increase with time, eventually attaining a steady - state value. 

\begin{figure}[htb]
\includegraphics[width=1.1\columnwidth, height=0.4\textheight]{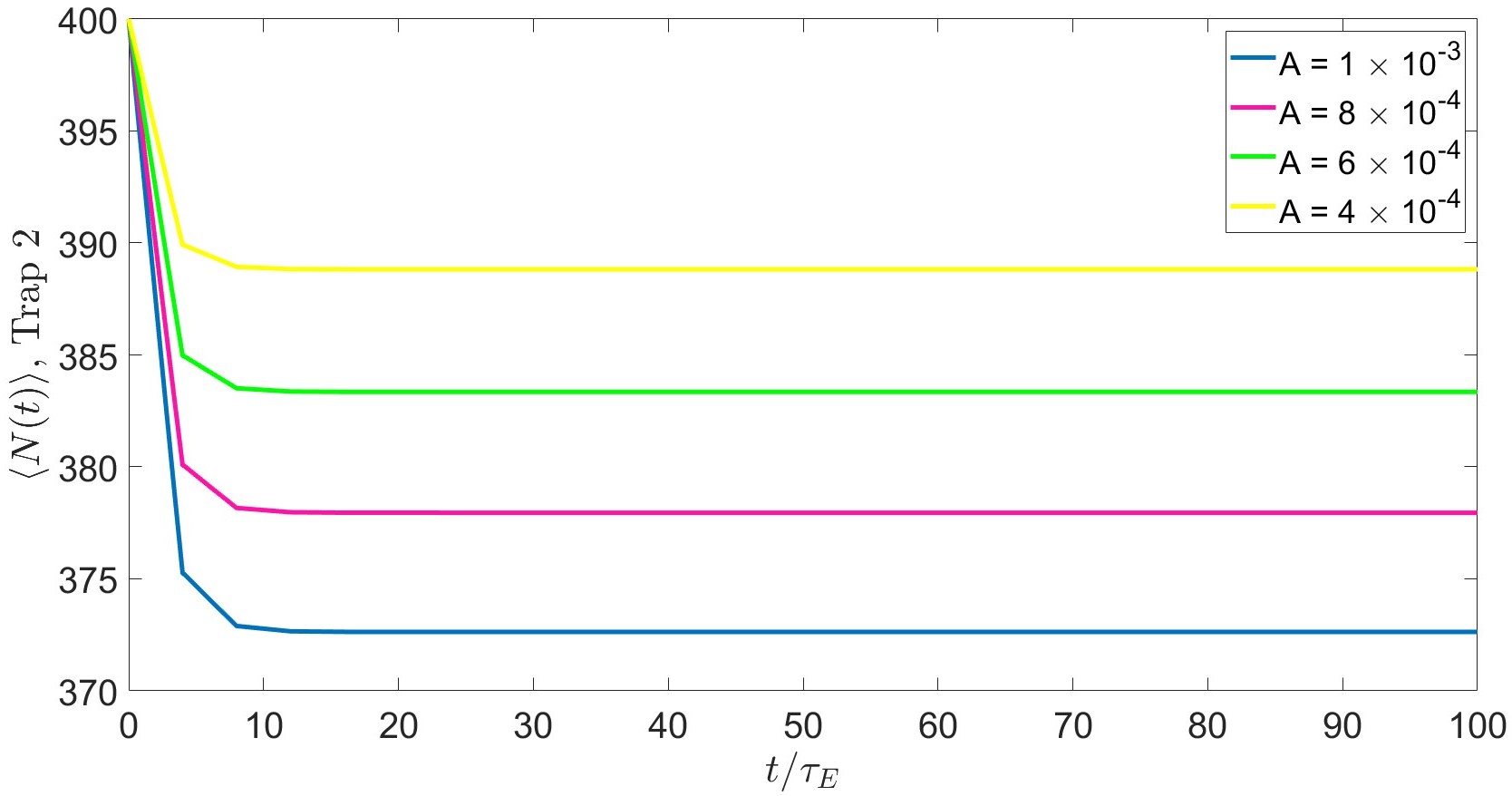}
\caption{\label{fig:numexpvalvaryA2} Time evolution of the expectation values of the particle number $N(t)$ in the second harmonic trap containing the ground state of the system, calculated with varying values of the coupling constant $A$ in the master equation, with these values of $A$ indicated in the figure.}
\end{figure}

Finally, let us now consider what happens when we vary the coupling constant $A$ for this case, which consequently varies the coupling strength between the trapped ultracold atom system and the background BEC. As can be seen in Fig. \ref{fig:numexpvalvaryA2}, lowering the value of the coupling constant $A$ will also lower the steady - state value of the expectation value of the particle number of the time - evolved component of the trapped ultracold atom state in trap 2, $\left\langle N_{2}(t)\right\rangle$. However, just as in the previous case, the manner in which the initial state in trap 2 evolves over time will not change. Hence, this implies that for the case wherein the expectation value of the particle number in trap 2 is greater than in the other two traps, weakening the coupling strength between the background BEC and the trapped ultracold atom gas will reduce the number of atoms in trap 2 that are lost as the state evolves over time, while ensuring that this state will still attain a steady - state over the course of its time evolution.

\subsection{$\left\langle N_{3}(t)\right\rangle > \left\langle N_{1}(t)\right\rangle$ and $\left\langle N_{3}(t)\right\rangle > \left\langle N_{2}(t)\right\rangle$}

\begin{figure}[htb]
\includegraphics[width=0.5\columnwidth, height=0.2\textheight]{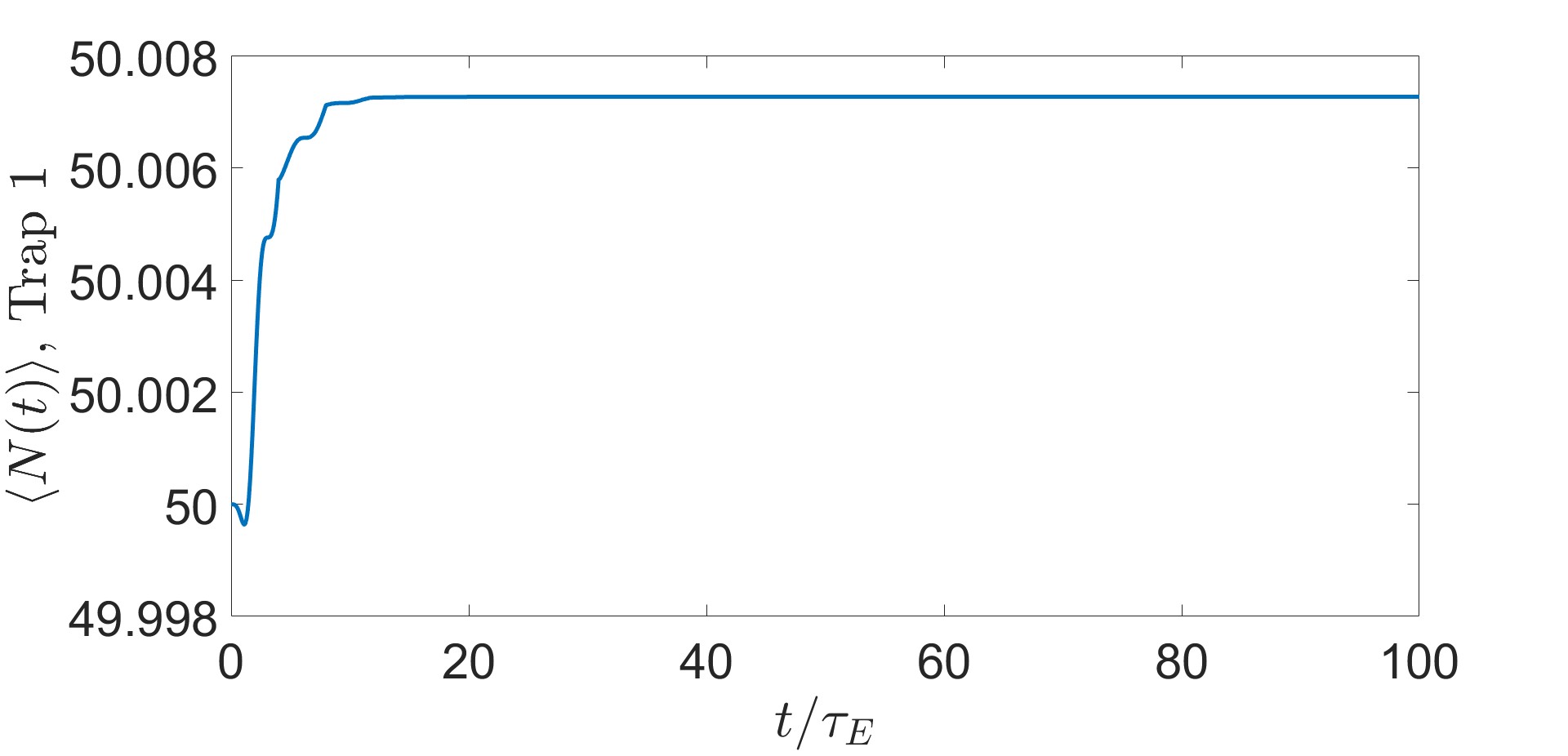}
\includegraphics[width=0.5\columnwidth, height=0.2\textheight]{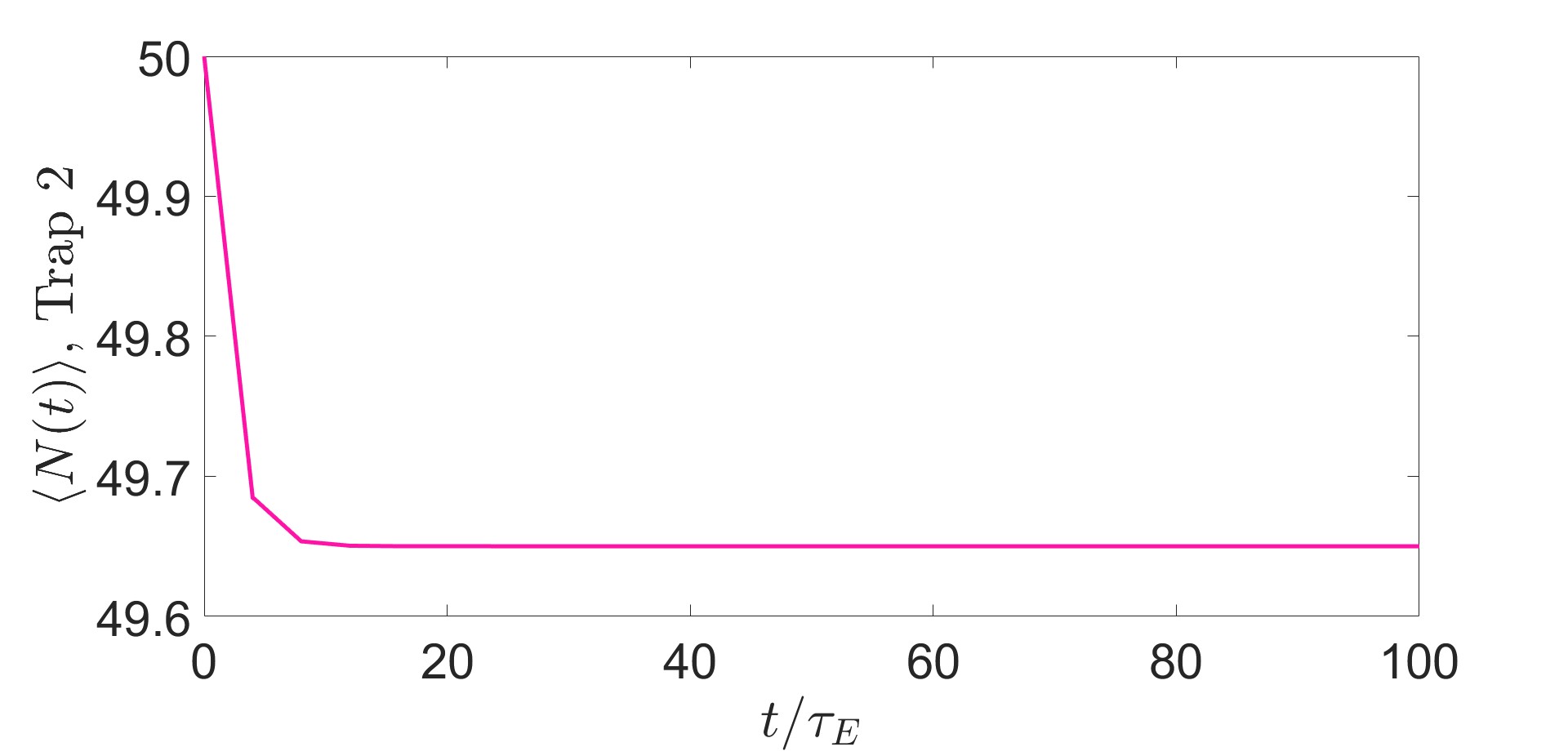}
\includegraphics[width=0.5\columnwidth, height=0.2\textheight]{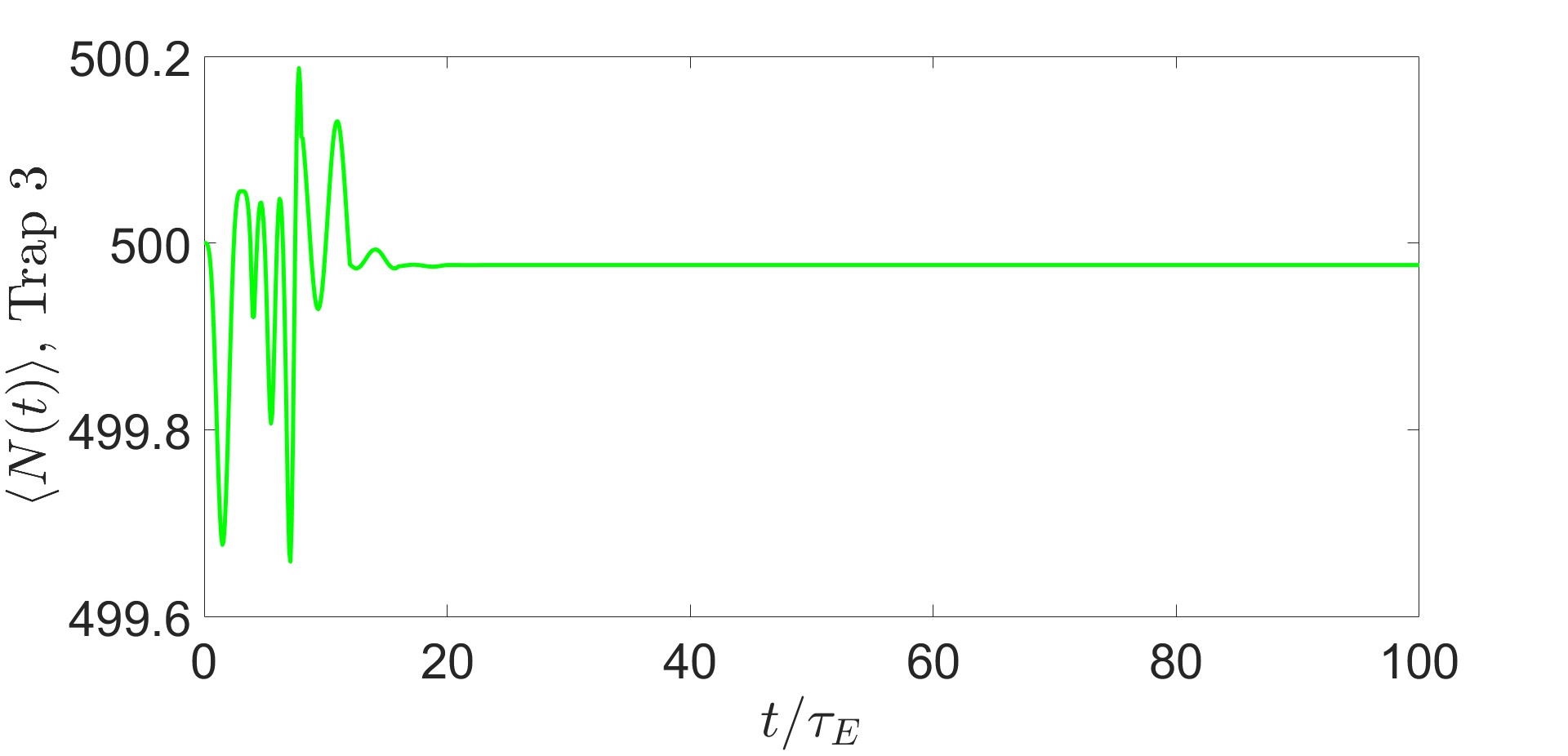}
\caption{\label{fig:numexpval8} Time evolution of the expectation values of the particle number $N(t)$ in the first (upper left), second (upper right) and third (lower left) harmonic traps containing the ground state of the system, calculated using a superposition of four particle number eigenstates per ground state location, with corresponding eigenvalues given in Eq. \ref{numopeigenvalcase8}.}
\end{figure}

Having considered the cases wherein $\left\langle N_{1}\right\rangle > \left\langle N_{2}\right\rangle and \left\langle N_{1}\right\rangle > \left\langle N_{3}\right\rangle$ as well as $\left\langle N_{2}\right\rangle > \left\langle N_{1}\right\rangle and \left\langle N_{2}\right\rangle > \left\langle N_{3}\right\rangle$, we now consider the case where $\left\langle N_{3}\right\rangle > \left\langle N_{1}\right\rangle and \left\langle N_{3}\right\rangle > \left\langle N_{2}\right\rangle$. Let us first look at how each component of the trapped ultracold atom gas in each of the harmonic potentials of the trap array evolve over time for this case. The particle number eigenvalues corresponding to the component particle number eigenstates that are used to construct the components of the initial state of the ultracold atom gas in traps 1, 2 and 3 of the harmonic trap array have the following values:
\begin{eqnarray}
&&(N_{g,1})_1 = 20, (N_{g,1})_2 = 80, (N_{g,1})_3 = 25, (N_{g,1})_4 = 75, (N_{g,2})_1 = 20, (N_{g,2})_2 = 80,\nonumber\\ 
&&(N_{g,2})_3 = 25, (N_{g,2})_4 = 75, (N_{g,3})_1 = 414, (N_{g,3})_2 = 586, (N_{g,3})_3 = 415, (N_{g,3})_4 = 585\nonumber\\
\label{numopeigenvalcase8} 
\end{eqnarray}
As can be seen in Fig. \ref{fig:numexpval8}, each of the trapped ultracold atom gas's components in traps 1, 2 and 3 will evolve over time in a manner similar to the previous two cases, which was shown in Figs. \ref{fig:numexpval2} and \ref{fig:numexpval4}. However, unlike the previous two cases, the change in $\left\langle N_{3}(t)\right\rangle$ at any instant of time is not large compared to the variation in $\left\langle N_{2}(t)\right\rangle$ and $\left\langle N_{1}(t)\right\rangle$. In fact, the oscillatory behavior for $\left\langle N_{3}(t)\right\rangle$ is still present for this case, just as it was present in the previous two cases, which is characteristic of the time evolution of components of the trapped ultracold atom gas in trap 3. Furthermore, for this case, the resulting steady - state value for $\left\langle N_{3}(t)\right\rangle$ is very close to, if not equal to, its initial value $\left\langle N_{3}\right\rangle$ despite fluctuating in value before settling to its steady - state value, which is the same behavior shown in the previous two cases for the time - evolved component of the trapped ultracold atom gas in trap 3.

\begin{figure}[htb]
\includegraphics[width=1.1\columnwidth, height=0.4\textheight]{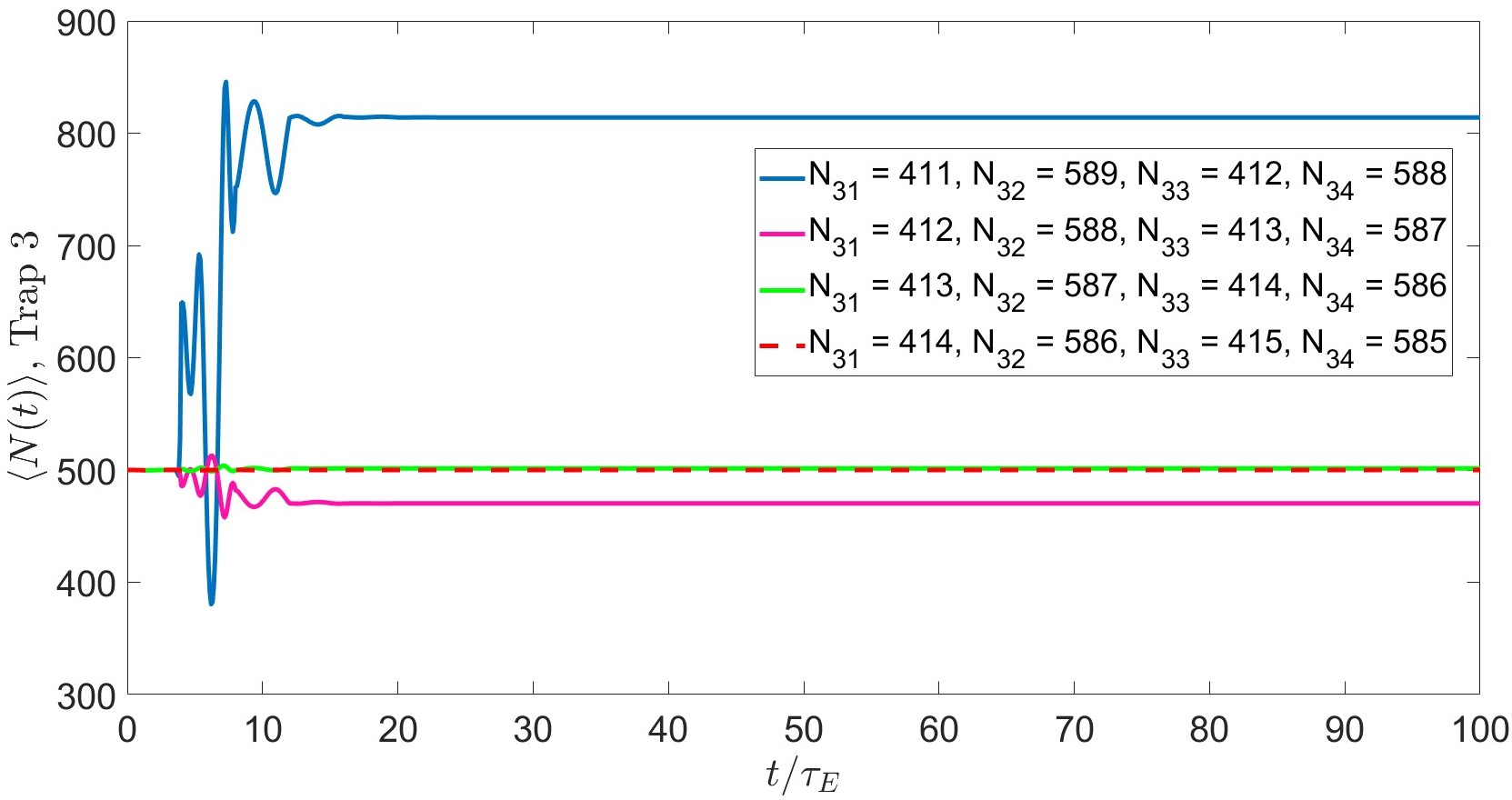}
\caption{\label{fig:numexpval6} Time evolution of the expectation values of the particle number $N(t)$ in the third  harmonic trap containing the ground state of the system, calculated using different superpositions of four particle number eigenstates per ground state location, with corresponding eigenvalues given in the figure.}
\end{figure}

Having seen how the trapped ultracold atom gas evolves over time in each of the harmonic potentials of the trap array for this case, let us now see what the effect of varying the component particle number eigenstates of the initial state corresponding to the component of the ultracold atom gas in Trap 3 will be on its time evolution and steady - state attained. As can be seen in Fig. \ref{fig:numexpval6}, varying the initial state of the system in Trap 3 will not change the manner by which the initial state of the ultracold atom gas's component in trap 3 evolves over time, but it will change the steady - state value of the particle number expectation value for this component of the trapped ultracold atom gas. In particular, from a steady-state value of $\left\langle N_{3}(t)\right\rangle \approx 800$ when the initial state in trap 3 has the form
\begin{equation}
\frac{1}{\sqrt{4}}\left(\left|(N_{g,3})_1 = 411\right\rangle+\left|(N_{g,3})_2 = 589\right\rangle+\left|(N_{g,3})_3 = 412\right\rangle+\left|(N_{g,3})_4 = 588\right\rangle\right),
\end{equation}
this steady - state value decreases to $\left\langle N_{3}(t)\right\rangle\approx 470$ when the initial state in trap 3 has the form
\begin{equation}
\frac{1}{\sqrt{4}}\left(\left|(N_{g,3})_1 = 412\right\rangle+\left|(N_{g,3})_2 = 588\right\rangle+\left|(N_{g,3})_3 = 413\right\rangle+\left|(N_{g,3})_4 = 587\right\rangle\right)
\end{equation}
Eventually, as the value of $(N_{g,3})_1$ and $(N_{g,3})_3$ increase, the steady - state value of $\left\langle N_{3}(t)\right\rangle$ will eventually approach a value of $\left\langle N_{3}(t)\right\rangle = 500$, which is the particle number expectation value of the initial state of the ultracold atom gas's component localized in trap 3 of the array. 

\begin{figure}[htb]
\includegraphics[width=1.1\columnwidth, height=0.4\textheight]{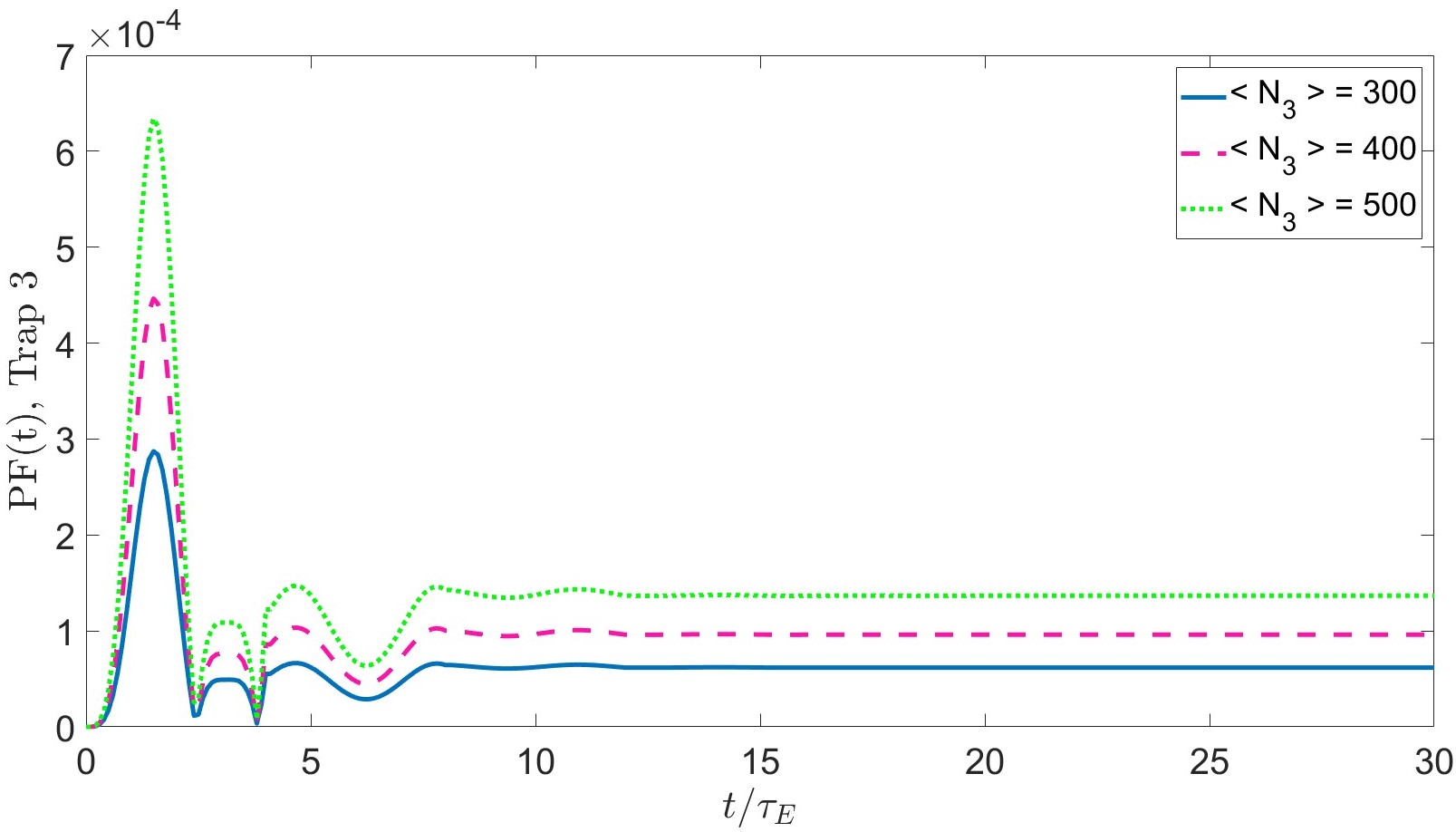}
\caption{\label{fig:condfrac3} Time evolution of the particle number fraction $PF(t)$ of the ultracold atom system's component in trap 3, with varying initial values of the particle number expectation value as indicated in the figure.}
\end{figure}

Having seen the effect of varying the components of the initial state of the component of the trapped ultracold atom gas in trap 3, let us now see what happens when we vary the particle number expectation value of the initial state of this component of the ultracold atom gas. Following our analysis of the effects of varying $\left\langle N_{1}\right\rangle$ and $\left\langle N_{2}\right\rangle$ when $\left\langle N_{1}\right\rangle > \left\langle N_{2}\right\rangle and \left\langle N_{1}\right\rangle > \left\langle N_{3}\right\rangle$ and $\left\langle N_{2}\right\rangle > \left\langle N_{1}\right\rangle and \left\langle N_{2}\right\rangle > \left\langle N_{3}\right\rangle$, respectively, we take a look at what happens to the particle fraction of the time - evolved component of the trapped ultracold atom gas in trap 3 for this case, with the particle fraction given by Eq. \ref{particlefraction}, with $j=3$. As shown in Fig. \ref{fig:condfrac3}, the manner by which $PF(t)$ evolves for this case will not change as $\left\langle N_{3}\right\rangle$ increases; rather, it is the maximum value and steady - state value of $PF(t)$ for the time - evolved component of the trapped ultracold atom gas in trap 3 that changes, both increasing with $\left\langle N_{3}\right\rangle$.

\begin{figure}[htb]
\includegraphics[width=1.1\columnwidth, height=0.4\textheight]{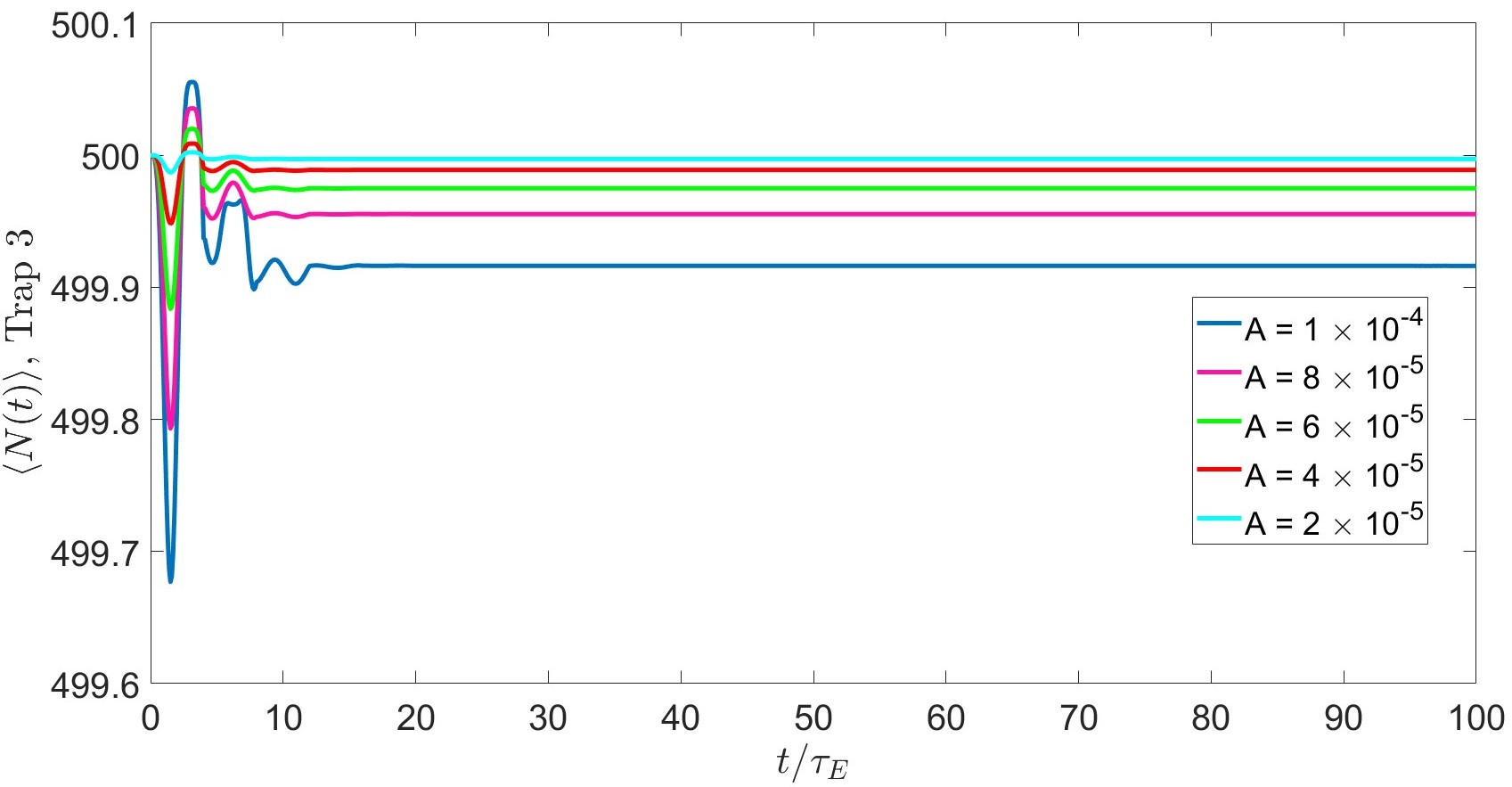}
\caption{\label{fig:numexpvalvaryA3} Time evolution of the expectation values of the particle number $N(t)$ in the third harmonic trap containing the ground state of the system, calculated with varying values of the coupling constant $A$ in the master equation, with these values of $A$ indicated in the figure.}
\end{figure}

Finally, let us examine the effect of weakening the coupling between the trapped ultracold atom system and the background BEC for this case by decreasing the magnitude of the coupling constant, $A$. As can be seen in Fig. \ref{fig:numexpvalvaryA3}, decreasing the coupling constant $A$'s magnitude will increase the magnitude of the steady - state value of the particle number expectation value $\left\langle N_{3}(t)\right\rangle$ of the time-evolved component of the trapped ultracold atom system in trap 3 for this case. At the same time, this is accompanied by a corresponding decrease in the maximum value of $\left\langle N_{3}(t)\right\rangle$ for this system. It is of interest to note that the manner in which the component of the trapped ultracold atom system in trap 3 evolves over time does not change even as the magnitude of the coupling constant $A$ changes. In particular, the instant of time when $\left\langle N_{3}(t)\right\rangle$ attains its maximum value does not, from Fig. \ref{fig:numexpvalvaryA3}, change even as $A$ changes, and the oscillations present in $\left\langle N_{3}(t)\right\rangle$ still occur at the same instants of time no matter the magnitude of $A$, though their amplitude will change as $A$'s magnitude changes. 

\section{Discussion}

Based on the results presented, we can see that, for an ultracold atom gas trapped in an array of 5 harmonic potentials arranged beside each other, which is coupled to a background BEC, it will evolve in such a way that if the trapped ultracold atom gas is trapped in 3 of these 5 potentials and occupy the ground state energies in these traps, and are in addition coupled to the excited state energies in the other two traps, then the combination of the driving of these ultracold atoms from the ground state to the excited state and the dissipation of Bogoliubov excitations into the background BEC by these driven ultracold atoms to enable them to return to the ground state energy will result in the ultracold atom gas evolving over time in such a way that the expectation value of the particle number of each component of the ultracold atom gas occupying one of the harmonic potential traps in the trap array will approach a steady - state value. At the same time, the manner by which the trapped ultracold atom gas evolves over time, as well as the particle number expectation value at the system's steady state, will depend on the particle number expectation value of the initial state of the system at each of the harmonic traps in the array, on the particle number eigenstates which form the initial state of the trapped ultracold atom system, and on the strength of coupling between the trapped ultracold atom system and the background BEC. 

However, what seems to be surprising among the results obtained is the distinctive manner by which the ultracold atom states in each harmonic trap in the array evolve towards their respective steady states. Notably, the results show that mixing between the atoms of the trapped ultracold atom gas and the background BEC can occur, increasing the expectation value of the particle number of the trapped ultracold atom gas, so long as the initial state of the system has more atoms in the first trap than in the other two. At the moment, the author cannot yet explain why indeed this is the case, and doing so is currently beyond the scope of this research work. The author would like to note that this behavior is a matter of interest emerging from this work, and being able to explain why this behavior in this system emerges in future research work may actually tell us something new about how two ultracold atom gases interact with each other. 

At the same time, the author would like to point out that similar behavior has been seen in another driven - dissipative system involving BECs which has been experimentally realized \cite{labouvie}. In this work, a weakly - interacting BEC is trapped in a one - dimensional periodic potential, with a roughly equal number of atoms occupying each site in the potential. The number of atoms in one of the sites is then reduced using scanning electron microscopy (SEM), ionizing the atoms and exciting them out of the potential site, thus serving as the source of dissipation in one of the potential sites. At the same time, tunneling between the sites allows for inter-site transport for the atoms, thus serving as the mechanism for driven dynamics in the system. By treating the periodic potential site undergoing SEM as our system, and the surrounding potential sites whose particle number remains the same as our superfluid reservoir, we find that, independent of the dissipation strength $\gamma$ (which is due to the intensity of the electron beam ionizing the atoms in the system site), the expectation value of the number of atoms in the system site will, due to the combined driven - dissipative dynamics, approach a steady state, with that steady state, and the manner by which the steady - state is approached, dependent on the number of particles initially in the system site. However, the system and environment considered in this paper, together with the driven - dissipative dynamics resulting in the emergence of a steady - state in the system, is markedly different from that considered by the authors of Ref. \cite{labouvie}. Nonetheless, these two results serve to underline the significance of using a BEC as an environment for open quantum systems used in the dissipative preparation of particular steady states, while at the same time presenting two different systems and two different environments which act as superfluid or excitation reservoirs, which can be used to achieve similar results via the driven - dissipative interaction between the system and the environment.

Nonetheless, the emergence of steady states with respect to the expectation value of the particle number for the trapped ultracold atom system undergoing driven - dissipative dynamics as described above implies that this system can possibly be used in the preparation of multiple ultracold atom states requiring a fixed particle number, in particular BECs. This dissipative quantum state preparation scheme becomes especially important when we consider that novel applications of BECs continue to emerge, such as for fundamental tests of quantum gravity \cite{howl, overstreet, thompson} and for quantum computing and quantum simulation \cite{byrnes, ghasemian, fujii,bello}. The need for BECs for these and other possible applications that are of interest in quantum mechanics and quantum technology, now and in the future, will ensure that dissipative quantum state preparation schemes for constant particle number states such as the one formulated in this paper, as well as other quantum state preparation schemes for these types of states, will continue to be of interest among theorists and experimentalists working in the field of ultracold atoms and quantum technology.

\section*{Acknowledgments}

The author would like to thank the Research Center for the Natural and Applied Sciences (RCNAS) and the College of Science of the University of Santo Tomas for financial support during the conduct of this research. Part of this work was conducted when the author was a visiting researcher at the Institute for Basic Science - Center for Theoretical Physics of Complex Systems (PCS - IBS), Daejeon, Republic of Korea. The author would also like to acknowledge S-H. Park, A. Patra, B-H. Kim, A. Parafilov, D. Safranek and S. Verma for useful suggestions towards the improvement of the manuscript. Finally, the author would like to acknowledge the invaluable support provided by Ms. T. B. O. Tejada.

\bibliographystyle{elsarticle-num-names}

\bibliography{DDHD_rev01}

\end{document}